\documentclass[aps, prd, amsmath, floats, floatfix, twocolumn,superscriptaddress, nofootinbib, showpacs]{revtex4-1}

\usepackage{graphicx}
\usepackage{amsmath,amssymb}
\usepackage{amsfonts}
\usepackage{xspace} 
\usepackage[usenames]{color}
\usepackage{dcolumn}
\usepackage{bm}
\usepackage{mathrsfs}
\usepackage{ulem}
\newcommand{\Cornell}{\affiliation{Center for Radiophysics and Space
    Research, Cornell University, Ithaca, New York, 14853}}
\newcommand{\Maryland}{\affiliation{Maryland Center for Fundamental
    Physics \& Joint Space-Science Institute, \\ Department of Physics, University of Maryland, College
    Park, MD 20742}}
\newcommand{\CITA}{\affiliation{Canadian Institute for Theoretical Astrophysics,
    University~of~Toronto, Toronto, Ontario M5S 3H8, Canada}}

\definecolor {darkgreen}{rgb}{0.2,0.7,0.2}

\begin{document}

\title{Reducing orbital eccentricity of precessing black-hole binaries}

\author{Alessandra Buonanno} \Maryland %
\author{Lawrence E. Kidder} \Cornell %
\author{Abdul H. Mrou\'e} \CITA%
\author{Harald P. Pfeiffer} \CITA %
\author{Andrea Taracchini} \Maryland %

\begin{abstract}
  Building initial conditions for generic binary black-hole evolutions
  which are not affected by initial spurious eccentricity remains a
  challenge for numerical-relativity simulations. This problem can be
  overcome by applying an eccentricity-removal procedure which
  consists of evolving the binary black hole for a couple of orbits,
  estimating the resulting eccentricity, and then restarting the
  simulation with corrected initial conditions.  The presence of spins
  can complicate this procedure. As predicted by post-Newtonian
  theory, spin-spin interactions and precession prevent the binary
  from moving along an adiabatic sequence of spherical orbits,
  inducing oscillations in the radial separation and in the orbital
  frequency. For single-spin binary black holes these
  oscillations are a direct consequence of monopole-quadrupole
  interactions. However, spin-induced oscillations occur at approximately twice
  the orbital frequency, and therefore can be distinguished and
  disentangled from the initial spurious eccentricity which occurs at
  approximately the orbital frequency.  Taking this into account, we
  develop a new eccentricity-removal procedure based on the derivative
  of the orbital frequency and find that it is rather successful in
  reducing the eccentricity measured in the orbital
  frequency to values less than $10^{-4}$ when
  moderate spins are present. We test this new procedure using
  numerical-relativity simulations of binary black holes with mass
  ratios $1.5$ and $3$, spin magnitude $0.5$, and various spin
  orientations. The numerical simulations exhibit
  spin-induced oscillations in the dynamics at approximately twice the
  orbital frequency. Oscillations of similar frequency are also visible in
  the gravitational-wave phase and frequency of the dominant $l=2$,
  $m=2$ mode.   
\end{abstract}

\date{\today \hspace{0.2truecm}}


\maketitle

\section{Introduction}
\label{sec:intro}
Over the last few years, numerical simulations of binary black
holes have improved tremendously (see e.g., the recent
reviews~\cite{Hannam:2009rd,Hinder:2010vn,Centrella:2010}).  These simulations are
now used to aid data analysts for gravitational-wave detectors in the
construction of analytical
templates~\cite{Buonanno2007,Ajith-Babak-Chen-etal:2007b,Damour2009a,
Buonanno:2009qa,Ajith2009,Pan:2009wj}, and in testing the efficiency
of detector pipelines by injecting numerical
waveforms~\cite{ninjashort}.

During the gravitational-radiation driven inspiral of a binary black
hole, the orbital eccentricity decreases very
quickly~\cite{PetersMathews1963,Peters1964}.  For binary black holes
formed from binary stellar evolution~\cite{Postnov:2007jv} (instead of
dynamical capture~\cite{Miller-Hamilton2002,Wen2003}), the
orbital eccentricity is expected to be essentially zero by the time
the binary enters the frequency band of ground-based
gravitational-wave detectors.  Therefore, it is important that
numerical simulations can be done for very low eccentricity binaries.

Performing black-hole simulations with very small orbital eccentricity
is not easy for several reasons.  Orbital parameters that result in
vanishing eccentricity are only known approximately through
post-Newtonian (PN) theory~\cite{Blanchet2006}. The translation of
orbital parameters from PN theory into a complete binary black-hole
initial-data set is ambiguous, because of differing coordinate systems
and effects arising from solving the non-linear Einstein constraint
equations~\cite{Pfeiffer2002a}.  And finally, early in a numerical
evolution each black hole relaxes toward a steady state, affecting the
black-hole masses, spins~\cite{DainEtAl:2008,Lovelace2008,Chu2009},
and orbital parameters.

The complete evolution of a binary black hole is determined by its
initial data.  Therefore, control of orbital eccentricity has to be
addressed in the construction of the initial data.  The first
approaches to construct low-eccentricity initial data were based on
the assumption of {\em circular} orbits with the orbital frequency
determined by the effective potential
method~\cite{cook94e,Baumgarte2000,pfeiffer-etal-2000} and the ``Komar
mass''
ansatz~\cite{Gourgoulhon2001,Grandclement2002,Tichy2003,Tichy2004,Caudill-etal:2006}. Both
methods were shown to give similar results~\cite{Caudill-etal:2006}.
Reference~\cite{Grigsby:2007fu} presented techniques to measure
eccentricity based on initial data alone.  When binary black-hole
evolutions became
possible~\cite{Pretorius2005a,Baker2006a,Campanelli2006a}, it was
realized that initial data constructed using the assumption of
circular orbits resulted in a {\em spurious} orbital eccentricity of
order one
percent~\cite{Buonanno-Cook-Pretorius:2007,Pfeiffer-Brown-etal:2007,Baker2006d},
primarily due to neglecting the radial inspiral velocity, and due to
the initial relaxation of the black holes.

Two techniques are in use to achieve an orbital eccentricity smaller
than what can be achieved with quasi-circular initial data. One
approach~\cite{Husa-Hannam-etal:2007} evolves PN equations for the
trajectories of the centers of the black holes.  This subsidiary
evolution of ordinary differential equations is started at large
initial separation, so that any spurious eccentricity due to the
initial conditions dies out and the binary settles down to an
inspiraling orbit with non-zero radial velocity. At the desired
separation, the subsidiary evolution is stopped, the positions and
velocities of the particles are read off, and are used as initial data
parameters for the construction of the initial data set for the
subsequent numerical evolution.  This approach reduces eccentricity to
about $0.002$ for equal-mass, non-spinning binaries, but is less
successful for unequal masses or high spins~\cite{Walther:2009ng}.

The second approach, proposed in Ref.~\cite{Pfeiffer-Brown-etal:2007}
and refined in Ref.~\cite{Boyle2007}, performs an iterative procedure 
(see also Ref.~\cite{Tichy:2010qa}).
One begins with initial data with reasonably low eccentricity, e.g.,
quasi-equilibrium initial data or initial data utilizing PN
information.  One evolves this initial data for about two to three
orbits, analyzes the orbit, and computes an improved initial data set
with (hopefully) lower eccentricity.  This procedure can be repeated
until the desired degree of eccentricity is obtained.

\begin{figure}[!ht]
  \begin{center}
    \includegraphics*[width=0.45\textwidth]{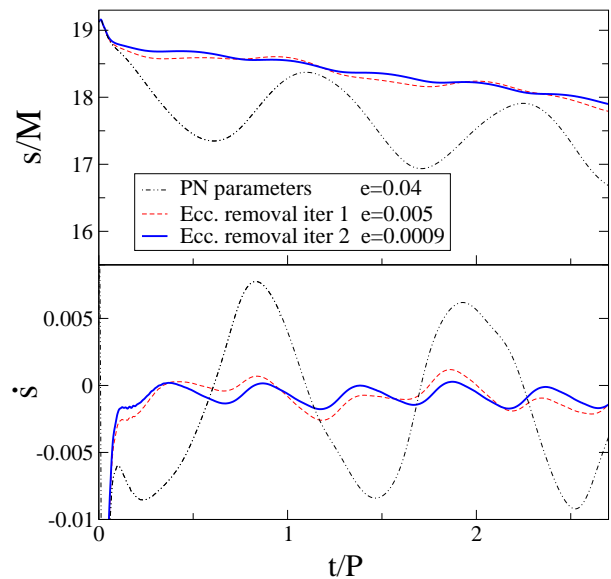}
    \caption{ \label{fig:EccRem-PropSep} Eccentricity removal based on
      {\bf proper separation} applied to a {\em precessing} binary
      black hole: The horizontal axis represents time in units of the
      initial orbital period $P=442M$.  For eccentricity $e\gtrsim
      0.01$, oscillations due to orbital eccentricity with period
      $\sim\! P$ dominate, and eccentricity removal is effective. For
      $e\sim 0.01$ oscillations at {\em one-half} the orbital period
      become apparent, spoiling further eccentricity removal based on
      $\dot{s}$.  In this example, the mass-ratio is $m_1/m_2=1.5$, the
      larger black hole carries spin $S_1=0.5 m_1^2$ with initial
      direction tangent to the orbital plane, and the smaller black
      hole has zero spin.  }
  \end{center}
\end{figure}

In past applications, eccentricity removal was based on the behavior
of the proper separation $s$ between the black hole apparent horizons.
For non-spinning black hole binaries~\cite{Buchman-etal-in-prep} or
binaries with spins parallel to the orbital angular
momentum~\cite{Chu2009,Lovelace:2010ne}, this works very well, and the eccentricity
drops by about an order of magnitude with each iteration.  However,
when one applies this eccentricity removal procedure to {\em
precessing} binaries, one encounters the difficulties illustrated in
Fig.~\ref{fig:EccRem-PropSep}.  At high eccentricity $e\gtrsim 0.01$,
$\dot{s}$ (where we indicate with a dot the derivative with
respect to time) shows the expected oscillations with a period
somewhat longer than the orbital period.\footnote{The period of
radial oscillations exceeds the orbital period because of periastron
advance~\cite{Mroue2010}.}  At sufficiently small eccentricity,
however, the proper separation $s$ and the radial velocity $\dot{s}$
exhibit oscillations at {\em twice} the orbital frequency (or
one-half the orbital period).  This frequency is distinct from the
frequency of oscillations caused by orbital eccentricity, and its
presence makes it very hard to further reduce
eccentricity based on an analysis of the proper separation $s$.

In this paper, we investigate these oscillations at twice the orbital
frequency, and develop techniques for eccentricity removal for
precessing binaries that mitigate the issues illustrated in
Fig.~\ref{fig:EccRem-PropSep}.  We can understand these oscillations
based on PN calculations.  In fact, as it was shown in
Refs.~\cite{Kidder:1995zr,Poisson:1997ha,Racine2008}, spin-spin
interactions in the dynamics and spin precession, can introduce
oscillations in the orbital separation and orbital frequency that
prevent the binary from moving along a sequence of spherical orbits.
Moreover, for single-spin binary black holes, the presence of
spin-induced oscillations in the dynamics is a direct consequence of
monopole-quadrupole interactions, that is of interactions of the form
$m_1\,S_2^2/m_2$ and $m_2\,S_1^2/m_1$.  It turns out that the
amplitude of the {\em spin-induced oscillations} in the orbital
frequency is half the amplitude of the oscillations in the coordinate
separation. Therefore, we propose to base the iterative eccentricity
removal on the orbital frequency and its time-derivative.  We develop
the relevant updating formulae for iterative eccentricity removal
based on the (time-derivative) of the orbital frequency, and
demonstrate with fully numerical simulations that iterative
eccentricity removal can proceed to much smaller eccentricities
$e\lesssim 10^{-4}$ that are measured in the orbital frequency.

We also find that PN theory predicts spin-induced oscillations in the 
separation with much smaller amplitude than
those visible in Fig.~\ref{fig:EccRem-PropSep}.  
Figure~\ref{fig:EccRem-PropSep} utilizes the {\em
proper separation $s$} between the apparent horizons along a line
element joining their centers.  We find that use of the {\em
coordinate separation} between the centers of
the apparent horizons instead results in much smaller
oscillations. Finally, we find that the spin-induced oscillations are
also present in the gravitational-wave frequency and phase, and are
qualitatively reproduced by the simple PN model used here.  

This paper is organized as follows. In Sec.~\ref{sec:ecc_AR}, we work
out the spin-induced oscillations in the radial separation and orbital
frequency for a PN model utilizing the Taylor-expanded PN Hamiltonian
with only the lowest order PN terms responsible for the spin-induced
oscillations.  We also compare the obtained analytic formulae with
numerical solutions of the ordinary differential equations describing
a PN binary.  In Sec.~\ref{sec:ecc_removal}, we present the new method
to reduce orbital eccentricity in presence of spins. In
Sec.~\ref{sec:ecc_NR}, we apply our improved eccentricity-removal
procedure to fully general-relativistic simulations of single and
double spin binary black holes, and compare the results with the
earlier eccentricity-removal procedure based on the proper horizon
separation. We also investigate the presence of spin-induced
oscillations in the gauge-invariant gravitational-wave phase and
frequency of a single-spin binary black hole. Finally, in
Sec.~\ref{sec:conclusions} we summarize our main conclusions.


\section{Eccentricity and spin-induced oscillations}
\label{sec:ecc_AR}

As mentioned in the introduction, spin-spin effects and
precession can induce oscillations in the orbital radial separation
and frequency preventing the binary black holes from moving along an
adiabatic sequence of spherical orbits. This result can be obtained in
a straightforward manner in PN
theory~\cite{Kidder:1995zr,Poisson:1997ha,Racine2008}.  Here we
re-analyse it in some detail using the PN Hamiltonian formalism.
 
As we shall see, when spins are not aligned with the initial direction
of the orbital angular momentum, the spin-induced oscillations are
unavoidable and their importance increases at smaller distances, since
at leading order spin-spin interactions scale as $1/r^3$, where $r$ is
the binary separation. Thus, if we were to start the binary black-hole
evolution at large separation with some initial orbital eccentricity,
we expect that, by the time the binary reaches smaller separations,
only spin-induced oscillations would be left.  However,
numerical-relativity simulations start the evolution at a separation
where the initial orbital eccentricity is not negligible. As we shall
discuss, there exists an efficient way to distinguish and disentangle
the initial orbital eccentricity from the spin-induced
oscillations, namely the typical frequency at which these two effects
occur.

Henceforth we use natural units $G=c=1$. 

\subsection{Eccentricity in Newtonian dynamics}
\label{Newtonian_dynamics}

Here we briefly review some useful formulae of Newtonian dynamics 
of eccentric orbits that we shall use below. 

In the center-of-mass frame, the two-body problem reduces to a
one-body problem for a particle of reduced mass $\mu = m_1\,m_2/M$,
subject to the acceleration $\ddot{\mathbf{r}} = - M/r^3\,\mathbf{r}$,
where $M = m_1 + m_2$ is the total mass of the binary. In the
Keplerian parametrization, a Newtonian orbit of eccentricity $e$ can
be described in terms of the eccentric anomaly $u$ (see, e.g.,
Ref.~\cite{Maggiore2008})
\begin{equation}
u(t)-e \sin{u(t)}=\bar{\Omega}\,t,\label{a
nom}
\end{equation}
where $\bar{\Omega} = \sqrt{M/a^3}$, $a$ being the semi-major axis,
so that
\begin{equation}
r(t)=\frac{\bar{r}}{1-e^2}\,\left[1-e\,\cos{u(t)}\right],\label{Newtr}
\end{equation}
where $\bar{r} = a\,(1-e^2)$. In the limit of small $e$ we can
approximate $r(t)$ using only the first harmonic, that is
\begin{equation}
r(t)=\bar{r}\left[1-e\,\cos{(\bar{\Omega}t)}\right]
+\mathcal{O}(e^2).\label{r(e)}
\end{equation}

In fact, although the frequency spectrum of $r(t)$ contains all
harmonics in $\bar{\Omega}$, a Fourier analysis of $r(t)$ shows that
harmonics beyond the first one are quite suppressed in presence of a
small eccentricity (see, e.g., Ref.~\cite{Maggiore2008}).  In the
Keplerian parametrization the orbital frequency reads
\begin{equation}
\Omega(t) =\frac{\bar{\Omega}\,\sqrt{1-e^2}}{\left[1-e\,\cos{u(t)}\right]^2}\,,
\end{equation}
and in the limit of small $e$ we find 
\begin{equation}
\Omega(t)= \bar{\Omega}\left [1+2e\,\cos{(\bar{\Omega}t)}\right ]
+\mathcal{O}(e^2).\label{Newtw}
\end{equation}
So, in Newtonian dynamics, whenever we have an eccentricity in $r(t)$,
we expect oscillations of amplitude $2e\,\bar{\Omega}$ at the
frequency $\bar{\Omega}$ in $\Omega(t)$.

\subsection{Two-body dynamics for spinning black holes in PN theory}
\label{sec:PNmodel}

We consider a binary composed of two black holes with masses $m_1$ and
$m_2$ and spins $\mathbf{S}_1$ and $\mathbf{S}_2$. The binary dynamics
can be described using the spinning Taylor-expanded PN Hamiltonian.  In the
center-of-mass frame, the Hamiltonian depends on the canonical
variables $(\mathbf{r},\mathbf{p})$ which describe the motion of a 
particle of reduced mass $\mu$, and on the spins $\mathbf{S}_1$
and $\mathbf{S}_2$.

For the purposes of our analysis, it is sufficient to restrict the
discussion to the Newtonian Hamiltonian, $H_\textrm{Newt}$, and
include only the leading 1.5PN spin-orbit (SO)
interaction~\cite{Damour-Schafer:1988}, $H_{\textrm{SO}}$, and the
leading 2PN spin-spin (SS) interaction~\cite{Barker:1975ae},
$H_{\textrm{SS}}$ , where the SS interaction includes
spin-induced monopole-quadrupole terms~\cite{Damour01c}. The Hamiltonian reads
\begin{equation}
H=H(\mathbf{r},\mathbf{p};\mathbf{S}_1,\mathbf{S}_2)=H_\textrm{Newt}+H_\textrm{SO}+H_\textrm{SS}\label{modelH},
\end{equation}
where 
\begin{align}
H_\textrm{Newt} &=\frac{\mathbf{p}^2}{2\mu}-\frac{\mu M}{r},\\
H_\textrm{SO}&=\frac{2}{r^3} \mathbf{S}_\textrm{eff} \cdot \mathbf{L},\\
H_\textrm{SS}&=\frac{\mu}{2Mr^3} \left[ 3 \left( \mathbf{S}_0 \cdot \mathbf{\hat{n}} \right)^2-\mathbf{S}_0^2 \right],
\end{align}
with $\mathbf{L} = \mathbf{r} \times \mathbf{p}$, $\mathbf{\hat{n}} = \mathbf{r}/|\mathbf{r}|$ and 
\begin{align}
\mathbf{S}_\textrm{eff}&=\left(1+\frac{3m_2}{4m_1} \right)\mathbf{S}_1+\left(1+\frac{3m_1}{4m_2} \right)\mathbf{S}_2,\\
\mathbf{S}_0&=\left(1+\frac{m_2}{m_1} \right)\mathbf{S}_1+\left(1+\frac{m_1}{m_2} \right)\mathbf{S}_2.
\end{align}
For reference, we point out that $\mathbf{S}_0$ can be rewritten in dimensionless form as 
\begin{equation}
\frac{\mathbf{S}_0}{M^2}= 
\frac{m_1}{M}\,\frac{\mathbf{S}_1}{m_1^2}
+\frac{m_2}{M}\,\frac{\mathbf{S}_2}{m_2^2}.
\end{equation}
The Hamilton equations of motion are given by
\begin{eqnarray}
\label{h1}
\dot{r}^i &=& \{r^i,H\}=\frac{\partial H}{\partial p_i},\\
\label{h2}
\dot{p_i} &=& \{p_i,H\}+F_i= -\frac{\partial H}{\partial r^i}+F_i,
\end{eqnarray}
where $F_i$ is the radiation-reaction force. Here we follow
Ref.~\cite{Buonanno06} and express $F_i$ in terms of the Newtonian
energy flux $\mathcal{F}_{\textrm{N}}={32} \mu^2/(5 M^2)\,v^{10}$ [see
Eqs.~(3.15), (3.27) in Ref.~\cite{Buonanno06})] where for
quasi-circular orbits $v= (M\,\Omega)^{1/3}$.  Equations (\ref{h1}),
(\ref{h2}) must be supplemented with the spin precession equations
\begin{eqnarray}
\label{S1}
\dot{S}^i_1 &=& \{S_1^i,H\}=\varepsilon^{ijk}\, \frac{\partial H}{\partial S^j_1}\,S_{1k} ,\\
\dot{S}^i_2 &=& \{S_2^i,H\}=\varepsilon^{ijk}\,\frac{\partial H}{\partial S^j_2}\,S_{2k},
\label{S2}
\end{eqnarray}
where $\varepsilon^{ijk}$ is the Levi-Civita symbol in flat spacetime.
%
%
The non-spinning conservative part of the dynamics together with the
lowest-order SO interactions allows the existence of spherical orbits
[$\mathit{r}(t)$=const.]~\cite{Damour01c}. In fact, if we consider
$H=H_{\rm{Newt}}+H_{\rm{SO}}$, it is straightforward to show that the
Hamiltonian is a spherically symmetric function that depends {\it
  only} on the radial separation and its conjugated momentum, i.e.,
$H=H(r,p_r)$. This happens because the other degrees of freedom are
constrained by the constants of motion: $\mathbf{L} ^2$ and
$\bf{S}_{\rm{eff}}\cdot \bf{L}$. More explicitly
\begin{equation}
H(r,p_r)=\frac{1}{2\mu}\left(p_r^2+\frac{\mathbf{L}^2}{r^2}\right)-
\frac{\mu M}{r}+\frac{2}{r^3}\bf{S}_{\rm{eff}}\cdot \bf{L}.
\end{equation} 
Imposing that at $t =0$, $r=r_0=\textrm{const.}$, we have
\begin{equation}
[\dot{r}]_0=\left [\frac{\partial H}{\partial p_r}\right ] _0
= \left [\frac{p_r}{\mu}\right ]_0=0
\end{equation}
and to have a stable spherical orbit we have to require also that
$[\dot{p}_r]_0=0$, hence
\begin{eqnarray}
  [\dot{p}_r]_0&=&\left [-\frac{\partial H}{\partial r}(r,p_r=0)\right]_0 
\nonumber \\
  &=& \left
    [\frac{\mathbf{L}^2}{\mu\,r^3}-\frac{\mu\,M}{r^2}
    +\frac{6}{r^4}\mathbf{S}_{\rm{eff}}\cdot
    \mathbf{L}\right ]_0=0. \label{SOsph}
\end{eqnarray}
Choosing properly $\mathbf{L}^2$ and $\bf{S}_{\rm{eff}}\cdot \bf{L}$
to satisfy Eq.~(\ref{SOsph}), we obtain spherical orbits. However,
once SS interactions are included, this is no longer true,
since $\mathbf{L}^2$ and $\bf{S}_{\rm{eff}}\cdot \bf{L}$ are no longer
constants of motion. Therefore, we must expect oscillations induced by SS terms 
in the radial separation and orbital frequency about their average values. 

\subsection{Oscillations induced by leading SS interactions: conservative dynamics}
\label{sec:ecc_conservative_SS}

In this section, we investigate the behavior of the radial separation
$r$ and of the orbital frequency $\Omega$ at 2PN order for the
conservative non-spinning dynamics. While doing so, we will also
assume a negligible precession of the spins and of the orbital plane,
since it takes place on a longer timescale than the effects we are
interested in.  Henceforth, we follow the method outlined in Appendix B 
of Ref.~\cite{Racine2008}.

As a first step, we restrict ourselves to the case in which
radiation-reaction is not present (i.e., $F_i=0$). As discussed
earlier, the presence of SS terms prevents $r$ and $\Omega$ from being
constant. Thus, we write~\cite{Racine2008}
\begin{equation}
r(t)=\bar{r}+\delta r(t),\quad 
\Omega(t)=\bar{\Omega}+\delta \Omega(t),\label{fluttua}
\end{equation}
where the bar stands for time-average $\langle \dots \rangle$ over one
orbital period; hence, by definition, $\langle \delta r(t)
\rangle=\langle \delta \Omega(t) \rangle=0$.  Our goal is 
to determine the equations that the oscillations $\delta r(t)$
and $\delta \Omega(t)$ must obey at 2PN order. For convenience, we
decompose vectors with respect to the triad defined by
\begin{equation}
  \mathbf{\hat{n}}=\frac{\mathbf{r}}{r}, 
\quad \mathbf{\hat{L}_\textrm{N}}=\frac{\mathbf{r}\times \mathbf{\dot{r}}}
  {\lvert \mathbf{r}\times \mathbf{\dot{r}} \rvert},
\quad  \boldsymbol{\hat{\lambda}}
= \mathbf{\hat{L}_\textrm{N}} \times \mathbf{\hat{n}}.
\end{equation}
This triad is such that $\mathbf{\hat{n}}$ and $\boldsymbol{\hat{\lambda}}$ are in the instantaneous orbital plane,
while $\mathbf{\hat{L}_\textrm{N}}$ is orthogonal to it.  In the
instantaneous orbital plane, we have the velocity
\begin{equation}
\mathbf{v} = \mathbf{\dot r}=\dot{r}\,\mathbf{\hat{n}} + \Omega\,r\,\boldsymbol{\hat{\lambda}}\, \label{velocity},
\end{equation}
and the acceleration 
\begin{equation}
\mathbf{a} =a_\textrm{rad}\,\mathbf{\hat{n}} + a_\textrm{tan}\,\boldsymbol{\hat{\lambda}}+a_{\perp}\mathbf{\hat{L}_\textrm{N}},
\end{equation}
with
\begin{equation}
a_\textrm{rad}=\mathbf{\hat{n}}\cdot\mathbf{\ddot{r}}=\ddot{r}-r\,\Omega^2,
\end{equation}
\begin{equation}
a_\textrm{tan}=\boldsymbol{\hat{\lambda}}
\cdot\mathbf{\ddot{r}}=\frac{1}{r}\frac{d}{dt}\left(r^2\,\Omega\right)\label{araddiff},
\end{equation}
and
\begin{equation}
  a_{\perp}=\mathbf{\hat{L}_\textrm{N}}\cdot\mathbf{\ddot{r}}=r\,\Omega\left(\boldsymbol{\hat{\lambda}}\cdot \frac{d\mathbf{\hat{L}_\textrm{N}}}{dt}\right).
\end{equation}
For future reference, we define the projection of $\mathbf{S}_0$ on the instantaneous orbital plane
\begin{equation}
\mathbf{S}_{0\perp}=\mathbf{S}_{0}-\mathbf{S}_{0} \cdot \mathbf{\hat{L}_\textrm{N}}.
\end{equation}
Note that Eq.~(\ref{velocity}) implicitly defines $\Omega$. We have
$\Omega=( \mathbf{\dot{r}} \cdot\boldsymbol{\hat{\lambda}}) /r$.
We need the acceleration $\mathbf{\ddot{r}}$ so we take a time derivative of Eq.~(\ref{h1}) and substitute
Eq.~(\ref{h2}) into that. We note that the Newtonian orbital 
angular momentum can be written as
\begin{equation}
\mathbf{L_\textrm{N}}=\mu\,\Omega\,r^2\,\mathbf{\hat{L}_\textrm{N}},
\end{equation}
while the orbital angular momentum $\mathbf{L} = \mathbf{r} \times
\mathbf{p}$ can be obtained from the Hamilton equation (\ref{h1}), that
is to say
\begin{equation}
\dot{\mathbf{r}}=\frac{\mathbf{p}}{\mu}+\frac{2}{r^3}\left(\mathbf{S}_{\textrm{eff}} \times \mathbf{r}\right),
\end{equation}
so that
\begin{equation}
\mathbf{L}=\mathbf{L_\textrm{N}}-\frac{2\mu}{r}\left[\mathbf{S}_\textrm{eff}- \mathbf{\hat{n}} \left( \mathbf{S}_\textrm{eff} \cdot \mathbf{\hat{n}} \right)\right] \label{LLN}.
\end{equation}
Since we want to work consistently at 2PN order, we replace $\mathbf{L}$ with Eq.~(\ref{LLN}) whenever it shows up in our calculations and drop higher PN terms. It is then straightforward to compute the radial component of the acceleration
\begin{eqnarray}
a_\textrm{rad}&=&-\frac{M}{r^2}\left\{1-\frac{2}{\mu M r^2}(\mathbf{S}_\textrm{eff} \cdot \mathbf{\hat{L}_\textrm{N}})\right.\nonumber\\
&&-\left.\frac{3}{2M^2r^2}\,\left[ 3 \left( \mathbf{S}_0 \cdot \mathbf{\hat{n}} \right)^2-\mathbf{S}_0^2 \right]\right\} \label{arad}
\end{eqnarray}
Since the leading-order spin acceleration is of 1.5PN order, we assume
that the radial oscillations scale at least as $\dot{r} = {\cal
  O}(3)$.\footnote{We denote the nPN order as ${\cal O}(2n)$} Hence, when computing the tangential component of the acceleration, at 2PN accuracy, we set $\dot{r} =0$ ($\mathbf{v} =
r\,\Omega\,\boldsymbol{\hat{\lambda}}$) and $p_r=0$ in every term coming from SO or SS interactions. Moreover, we 
also neglect any term depending on the time derivative of the spin in $a_{\textrm{tan}}$, since it is of higher PN order. This means that we are implicitly assuming that the spins are
constant. At 2PN order, we are left with
\begin{equation}
a_\textrm{tan}= 
-\frac{3}{M\,r^4} \left(\mathbf{S}_0 \cdot \mathbf{\hat{n}}\right)( \mathbf{S}_0 \cdot \boldsymbol{\hat{\lambda}}).\label{atan}
\end{equation}
%
%
Combining Eq.~(\ref{araddiff}) with
Eq.~(\ref{atan}), we solve for $r^2\,\Omega$ by integrating
Eq.~(\ref{atan}).  In doing that, we assume that $r$ and $\Omega$
in the right-hand side of Eq.~(\ref{atan}) are constants, as their
time derivatives are at least ${\cal O}(3)$, and also the spins are
constants, that is they do not precess.  Thus, under those
assumptions, the time evolution of the triad $\{\mathbf{\hat{n}},\
\boldsymbol{\hat{\lambda}},\ \mathbf{\hat{L}_\textrm{N}}\}$ is such
that $\mathbf{\hat{n}}$ and $\boldsymbol{\hat{\lambda}}$ swipe the
orbital plane at a frequency $\bar{\Omega}$ while
$\mathbf{\hat{L}_\textrm{N}}$ stays fixed:
\begin{eqnarray}
\label{evn}
\mathbf{\hat{n}}(t)&=&\cos(\bar{\Omega}t)\,\mathbf{\hat{n}}_0+
\sin(\bar{\Omega}t)\,\boldsymbol{\hat{\lambda}}_0, \\
\boldsymbol{\hat{\lambda}}(t)&=&-\sin(\bar{\Omega}t)\,\mathbf{\hat{n}}_0
+\cos(\bar{\Omega}t)\,\boldsymbol{\hat{\lambda}}_0,\\
\mathbf{\hat{L}_\textrm{N}}(t)&=&\mathbf{\hat{L}_{\rm N\,0}},
\label{evL}
\end{eqnarray}
where $\mathbf{\hat{n}}_0=\mathbf{\hat{n}}(0)$,
$\boldsymbol{\hat{\lambda}}_0=\boldsymbol{\hat{\lambda}}(0)$ and
$\mathbf{\hat{L}_\textrm{N\,0}}=\mathbf{\hat{L}_\textrm{N}}(0)$, and also
\begin{equation}
 \mathbf{\dot{\mathbf{\hat{n}}}}=\bar{\Omega}\,\boldsymbol{\hat{\lambda}}\,
\quad \quad \quad\boldsymbol{\dot{\mathbf{\hat{\lambda}}}}=-\bar{\Omega}\,\mathbf{\hat{n}}.
\end{equation}
Moreover, since we assume that the spins remain constant, we formally set 
$\mathbf{S}_1(t)=\mathbf{S}_1(0)$ and
$\mathbf{S}_2(t)=\mathbf{S}_2(0)$, so in what follows $\mathbf{S}_0=
\mathbf{S}_0 (0)$ and $\mathbf{S}_\textrm{eff}=\mathbf{S}_\textrm{eff}
(0)$.
By inserting Eqs.~(\ref{fluttua}) into Eqs.~(\ref{arad}) and
(\ref{atan}), one obtains a pair of coupled
differential equations:
%
%
\begin{eqnarray}
&&\delta\ddot{r}-\bar{r}\,\bar{\Omega}^2-\bar{\Omega}^2\,\delta r-2\bar{r}\,\bar{\Omega}\,\delta\Omega=-
\frac{M}{\bar{r}^2}\,\times\nonumber \\ 
&&\times \left \{1-\frac{2\bar{\Omega}}{M}(\mathbf{S}_\textrm{eff} \cdot \mathbf{\hat{L}_\textrm{N}})\right.\,
\nonumber\\
&& \left. -\frac{3}{2M^2\,\bar{r}^2} \left[ 3 \left( \mathbf{S}_0 \cdot \mathbf{\hat{n}} \right)^2-\mathbf{S}_0^2\right]\right\}+2\frac{M}{\bar{r}^3}\delta r \label{diff1}
\end{eqnarray}
and
\begin{eqnarray}
2 \bar{\Omega}\,\bar{r}\,\delta r+\bar{r}^2\,\delta\Omega=k-\frac{3}{2M\,\bar{\Omega}\,\bar{r}^3}( \mathbf{S}_0\cdot \mathbf{\hat{n}})^2.\label{diff2}
\end{eqnarray}
Here, $k$ is an integration constant  and, again, in the
right-hand-side of the above equations we keep only terms through 2PN order.  
To fix $k$ we time-average the above equations. We have
\begin{multline}
\langle  \left(\mathbf{S}_\textrm{0} \cdot \mathbf{\hat{n}}(t) \right)^2\rangle=S_{0i}\,S_{0j}\,\langle n^i(t)\,n^j(t)\rangle= \\
=S_{0i}\,S_{0j} \frac{1}{2}\left(\delta^{ij}-\hat{L}^i_{N0}\,\hat{L}^j_{N0}\right)=\frac{1}{2}\left[\mathbf{S}_0^2-( \mathbf{S}_0 \cdot  \mathbf{\hat{L}_\textrm{N\,0}})^2 \right]\,,
\end{multline}
obtaining
\begin{equation}
k=\frac{3}{4M\,\bar{\Omega}\,\bar{r}^3}\,\left[\mathbf{S}_0^2-( \mathbf{S}_0 \cdot  \mathbf{\hat{L}_\textrm{N\,0}} )^2 \right].
\end{equation}

Taking the time average of Eq.~(\ref{diff1}), we derive 
the following modified version of Kepler's law relating $\bar{r}$ and $\bar{\Omega}$
\begin{equation}
\bar{\Omega}^2=\frac{M}{\bar{r}^3}-\frac{2 \bar{\Omega}}{\bar{r}^3}\mathbf{S}_\textrm{eff} \cdot \mathbf{\hat{L}_\textrm{N\,0}}+\frac{3}{4M\bar{r}^5}\left[ 3 ( \mathbf{S}_0 \cdot \mathbf{\hat{L}_\textrm{N\,0}} )^2-\mathbf{S}_0^2\right].\label{Kepler}
\end{equation}
We decouple Eq.~(\ref{diff1}) from Eq.~(\ref{diff2}), then we use Eq.~(\ref{Kepler}). Since 
we expect that $\delta r=\mathcal{O}(4)$, we find that at 2PN order
\begin{equation}
\delta\ddot{r}+\bar{\Omega}^2\,\delta r =-\frac{3}{4M\,\bar{r}^4}\left[( \mathbf{S}_0 \cdot \boldsymbol{\hat{\lambda}} )^2-( \mathbf{S}_0 \cdot  \mathbf{\hat{n}})^2 \right],\label{diffdr}
\end{equation}
in agreement with Eq.~(B13) of Ref.~\cite{Racine2008}. The solution of the homogeneous equation is simply 
\begin{equation}
\delta r(t)_\textrm{hom}=A_r\cos{(\bar{\Omega}t+\varphi_r)}\,, \label{hom}
\end{equation}
where $A_r$ and $\varphi_r$ are fixed by the initial
conditions. Equation (\ref{hom}) describes possible oscillations due
to the initial eccentricity of the orbit. This eccentricity occurs at
the average orbital frequency and in principle can be removed as long
as quasi-circular initial conditions are enforced. Note that
Eq.~(\ref{hom}) is also consistent with the Newtonian result of
Eq.~(\ref{r(e)}).\footnote{Note that Eq.~(\ref{r(e)}) does not depend
  on $\varphi$ because in Eq.~(\ref{Newtr}) the radial separation at
  $t=0$ is picked to be equal to the semi-major axis.}  It is worth
noting that Eq.~(\ref{r(e)}) was derived as an expansion for a small
eccentricity $e$, while in this section we never explicitly referred
to $e$ at all. As a matter of fact, we are dealing with quasi-circular
orbits here, so that consistency between Eqs.~(\ref{r(e)}) and
(\ref{hom}) should be expected.

On the other hand, the spin-induced oscillations are described by the
particular solution of Eq.~(\ref{diffdr}) which reads
\begin{equation}
\delta r_\textrm{part}(t)=\frac{1}{4M^2\,\bar{r}}\left[( \mathbf{S}_0 
\cdot \boldsymbol{\hat{\lambda}}(t) )^2 -(\mathbf{S}_0 \cdot \mathbf{\hat{n}} (t))^2\right].\label{partr}
\end{equation}
These oscillations are a signature of SS interactions since they
depend on $\mathbf{S}_0$, which enters $H_\textrm{SS}$.  Once we know
$\delta r=\delta r_\textrm{hom}+\delta r_\textrm{part}$, we solve
Eq.~(\ref{diff2}) for $\delta\Omega$. Similarly to what we found above,
the homogeneous solution accounts for the initial conditions, while 
the particular solution accounts for the oscillations induced by SS effects. 
\begin{equation}
\delta \Omega_\textrm{part}(t)=\frac{\bar{\Omega}}{4M^2\,\bar{r}^2}\left[( \mathbf{S}_0 
\cdot \boldsymbol{\hat{\lambda}}(t) )^2 -(\mathbf{S}_0 \cdot \mathbf{\hat{n}} (t))^2\right].\label{partomega}
\end{equation}
The above equation is also consistent with the Newtonian result of 
Eq.~(\ref{Newtw}).

So far, we have assumed the nonspinning dynamics to be Newtonian. If we 
included nonspinning PN corrections through 3PN order in the Hamiltonian, 
we would still find the particular solutions (\ref{partr}) and (\ref{partomega}), 
but in this case the oscillations would occur at a frequency which will differ 
from Eq.~(\ref{Kepler}) because of nonspinning PN corrections.

Using the previous results, it is straightforward 
to compute the time derivatives of the oscillations $\delta \dot{r}$, 
Eq.~(\ref{partr}), and $\delta \dot{\Omega}$, Eq.~(\ref{partomega}). 
They read
\begin{eqnarray}
\delta\dot{r}(t)  &=&B_r\,\sin{(\bar{\Omega}t+\varphi_r)}\nonumber\\
&&-\frac{\bar{\Omega}}{M^2\,\bar{r}}\,\left( \mathbf{S}_0 \cdot \mathbf{\hat{n}} (t)\right)\,
( \mathbf{S}_0 \cdot \boldsymbol{\hat{\lambda}} (t)),\label{deltadotr}\\
\delta\dot{\Omega}(t) &=&B_{\Omega}\,\sin{(\bar{\Omega}t+\varphi_{\Omega})}\nonumber\\
&&-\frac{\bar{\Omega}^2}{M^2\,\bar{r}^2}\,\left( \mathbf{S}_0 \cdot \mathbf{\hat{n}} (t)\right)\,
( \mathbf{S}_0 \cdot \boldsymbol{\hat{\lambda}} (t)).\label{deltadotomega}
\end{eqnarray}
We note that when the spins are initially aligned or antialigned to
$\mathbf{\hat{L}_\textrm{N\,0}},$ the SS oscillations disappear, since
in this situation $\mathbf{S}_0$ remains perpendicular to $\mathbf{\hat{n}}$ 
and $\boldsymbol{\hat{\lambda}}$ throughout the evolution.
We see that for both quantities the time dependence of the SS term is
\begin{equation}
\left( \mathbf{S}_0 \cdot \mathbf{\hat{n}} (t)\right)\,( \mathbf{S}_0 
\cdot \boldsymbol{\hat{\lambda}} (t))=C\,\sin{(2\bar{\Omega}\,t+\gamma)},\label{twice}
\end{equation}
where 
\begin{equation}
C =\frac{(\mathbf{S}_0 \cdot \mathbf{\hat{n}}_0)^2+ 
(\mathbf{S}_0 \cdot \boldsymbol{\hat{\lambda}}_0)^2}{2} \label{C}
\end{equation}
and $\gamma$ satisfies
\begin{eqnarray}
\sin\gamma = \frac{1}{2}\sin(2\alpha), \nonumber \\
\cos\gamma = -\frac{1}{2}\cos(2\alpha),\label{gamma}
\end{eqnarray}
with $\cos\alpha = \mathbf{\hat S}_{0\perp} \cdot \mathbf{\hat{n}}_0$.
Thus, the spin-induced oscillations occur at {\it twice} the average
orbital frequency, and they can be neatly disentangled from the
eccentricity induced by initial conditions which occurs at the average
orbital frequency.  

Moreover, the amplitude of spin-induced oscillations is quite
  small.  To place their amplitude into perspective, consider a binary
  with orbital eccentricity $e$.  Taking a time-derivative of
  Eqs.~(\ref{Newtr}) and~(\ref{Newtw}), and comparing to
  Eqs.~(\ref{deltadotr}) and~(\ref{deltadotomega}) we find $B_r=\bar
  r\bar\Omega\, e$ and $B_\Omega=2\bar\Omega^2\,e$.  Equating the
  amplitudes of the spin-induced oscillations with the amplitude of
  the eccentricity-induced term, we see that spin-induced oscillations
  dominate only for eccentricities 
\begin{equation}\label{eq:spin-induced--ecc-bound}
     e < \left\{\begin{aligned}
& \frac{1}{2}\frac{S_{0\,\perp}^2}{M^4}\left(\frac{\bar r}{M}\right)^{-2}
\quad\mbox{for $\delta \dot r$,}\\
& \frac{1}{4}\frac{S_{0\,\perp}^2}{M^4}\left(\frac{\bar r}{M}\right)^{-2}
\quad\mbox{for $\delta \dot \Omega$.}\\
\end{aligned}
\right.
  \end{equation}
Numerical binary black-hole simulations typically start at a separation
    $\bar r/M\approx 15$, and in that case, spin-induced oscillations
    will dominate $\delta \dot r$ and $\delta \dot\Omega$ only for
    $e<0.002 S_{0\,\perp}^2/M^4$ and for $e<0.001S_{0\,\perp}^2/M^4$,
    respectively.  For maximally spinning black holes with
    least-favorable spin orientations, $S_{0\,\perp}/M^2=1$, so that
    even in this case spin-induced oscillations become relevant only
    at eccentricities of $\lesssim 0.001$.  For smaller spins, their
    effect is still smaller.  We note that spin-induced oscillations
    do affect $\delta\dot\Omega$ somewhat less than $\delta\dot r$,
    indicating that eccentricity-removal based on the orbital
    frequency will be preferable.

Let us notice that were we including the precession of the spins, 
the characteristic frequency at which the spin-induced oscillations occur would change. 
This can easily be seen if we assume that the precession is mainly due to SO 
effects, with $\mathbf{S}_1$ and $\mathbf{S}_2$ precessing about
$\mathbf{\hat{L}_{\textrm{N}}}$ at frequencies $\Omega_1$ and
$\Omega_2$. In this case, using Eqs.~(\ref{S1}), (\ref{S2}), we derive
\begin{eqnarray}
\Omega_1&=&\frac{2\mu\,\bar{\Omega}}{\bar{r}}\,\left(1+\frac{3\,m_2}{4\,m_1}\right),\\
\Omega_2&=&\frac{2\mu\,\bar{\Omega}}{\bar{r}}\,\left(1+\frac{3\,m_1}{4\,m_2}\right).
\end{eqnarray} 
If in Eqs.~(\ref{twice}) (\ref{C}) and (\ref{gamma}), we let $\mathbf{S}_0$ precess, 
we obtain that oscillations occur at frequencies given by linear combinations of
$\bar{\Omega}$, $\Omega_1$ and $\Omega_2$, namely 
$2\bar{\Omega}-\Omega_1-\Omega_2$, $2(\bar{\Omega}-\Omega_1)$ and
$2(\bar{\Omega}-\Omega_2)$. For the binary black-hole evolutions considered in 
this paper, $\Omega_{1,2} \ll \bar\Omega$, so the spin-induced oscillations occur 
at the frequency $2\bar{\Omega}$.

Lastly, the results of this section could be extended to higher PN orders by including 
next-to-leading SO terms (2.5PN order~\cite{Damour:2007nc,Porto:2010tr,Perrodin:2010dy,Levi:2010zu}) 
and SS terms (3PN order~\cite{SHS07,SSH08,SHS08,hergt_schafer_08,PR08a,PR08b}).
 
\subsection{Oscillations induced by leading SS terms: inspiraling dynamics}
\label{sec:ecc_RR}

In this section we compare the approximate analytical predictions for
$\delta \dot{r}$ and $\delta \dot{\Omega}$ with the results obtained
by numerically integrating the Hamilton equations (\ref{h1}) and
(\ref{h2}), including the radiation reaction force $F_i$. 
Since we actually want to extract the spin-induced
 oscillations, we need to remove the homogeneous part which is
 due to the eccentricity introduced by the initial conditions. In
 fact, in the presence of radiation-reaction the initial radial
 velocity has to be carefully chosen to guarantee that the binary moves
 along a quasi-adiabatic sequence of spherical orbits, progressively
 shrinking. Those initial conditions have been worked out in the
 analytical PN dynamics at post-circular~\cite{Buonanno00} and
 post-post-circular~\cite{DN2007b} orders. However, those initial
 conditions become more and more approximate if we start the
 evolution of the Hamilton equations at smaller and smaller
 separations. Moreover, from the study of the conservative dynamics in
 Sec.~\ref{sec:ecc_conservative_SS}, we understood that because of SS
 interactions it is impossible to have spherical orbits. 
To remove the oscillations at a frequency $\bar{\Omega}$ (the homogeneous part), we perform 
a fit of the data with a function 
 \begin{equation}
 f_\textrm{oscill}(t;B_\textrm{fit},\,\omega_\textrm{fit},\,\varphi_\textrm{fit})=B_\textrm{fit}\,
\sin(\omega_\textrm{fit}t+\varphi_\textrm{fit});
 \end{equation}
 where $\omega_\textrm{fit}$ is close to $\bar{\Omega}$. We subtract
 the fitted function $f_\textrm{oscill}$ from the raw data sample,
 obtaining a residual that oscillates at a frequency $2\bar{\Omega}$,
 superimposed to the smooth numerical inspiral. The reason why we need
 to fit also the frequency $\omega_\textrm{fit}$ is that there is an
 ambiguity as to what value we use for $\bar{\Omega}$. In
 principle, this value should be the average orbital frequency, but a
 priori we can only use the initial value $\Omega_0=\Omega(0)$ because
 we do not have an analytic prediction for $\bar{\Omega}$.  This is
 also true for the value of $\bar{r}$, which we replace with
 $r_0=r(0)$ throughout. We want to compare these residuals with
 analytical predictions based on Eqs.~(\ref{deltadotr})
 and~(\ref{deltadotomega}). We use the expression of the Newtonian
 flux to derive the effect of the radiation reaction (RR) on the two
 quantities $\dot r$ and $\dot \Omega$. Within the context of
 Newtonian dynamics, we have
\begin{equation}
\dot{r}_{\textrm{RR}}(t)=-\frac{64}{5}\,\mu\, M^2\,\left(r_0^4-\frac{256}{5} \mu\, M^2\, t\right)^{-3/4},
\end{equation}
where $r_0=r(0)$ and
\begin{equation}
r_{\textrm{RR}}(t)=\left(r_0^4-\frac{256}{5} \mu\, M^2\, t\right)^{1/4}.
\end{equation} 
A similar expression can be found also for $\dot{\Omega}_{\textrm{RR}}$. Considering that in a quasi-circular inspiral
\begin{equation}
\frac{\dot{r}_{\textrm{RR}}}{r_{\textrm{RR}}}=-\frac{2}{3}\frac{\dot{\Omega}_{\textrm{RR}}}{\Omega_{\textrm{RR}}}
\end{equation}
and $r^3_{\textrm{RR}}\Omega^2_{\textrm{RR}}=M$, we find
\begin{equation}
\dot{\Omega}_{\textrm{RR}}(t)=-\frac{3}{2}M^{1/2}\,\frac{\dot{r}_{\textrm{RR}}(t)}{r^{5/2}_{\textrm{RR}}(t)}.
\end{equation}
Therefore our analytical predictions will be given by
\begin{eqnarray}
\dot r_\textrm{pred}(t)&=&\dot r_\textrm{RR}(t)+\delta \dot r_\textrm{part}(t) \label{rdotpred}\\
\dot \Omega_\textrm{pred}(t)&=&\dot \Omega_\textrm{RR}(t)+\delta \dot \Omega_\textrm{part}(t),\label{omegadotpred}
\end{eqnarray}
where $\delta \dot r_\textrm{part}(t)$ and $\delta \dot
\Omega_\textrm{part}$ are given by the second term in the RHS of
Eqs.~(\ref{deltadotr}) and (\ref{deltadotomega}).

\begin{figure}[!ht]
\begin{center}
\includegraphics*[width=0.45\textwidth]{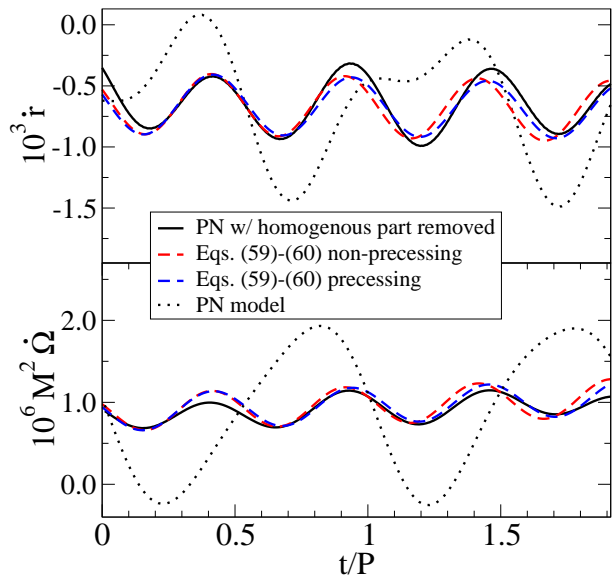}
\caption{{\bf Spin induced oscillations in PN model.}  We
    compare two PN calculations of spin-induced oscillations. The
    dashed lines are the predictions from Eqs.~(\ref{rdotpred})
    and~(\ref{omegadotpred}), with the two lines differing by whether
    $\mathbf{\hat{L}_\textrm{N}}$ is held constant
    (``non-precessing'') or evolving (``precessing'').  The solid line
    represents a solution of the full PN equations of motions, with
    the homogenous oscillations fitted and subtracted.  (Here, mass-ratio
    $q=2$, with maximal spins of initial orientations
    $(\theta_1=\pi/3,\phi_1=0)$ and $(\theta_2=2\pi/3,\phi_2=\pi/3)$
    at the initial orbital frequency $M \Omega_0=0.015$, that is an average initial
     orbital period $P=418M$, and the dotted
    curve represents the solutions of the PN model including inspiral
    motion). 
  \label{PNprediction}}
\par\end{center}
\end{figure}

Figure~\ref{PNprediction} shows the results for a
particular black-hole binary with mass ratio $q = m_2/m_1 = 2$ and
maximal spin magnitudes $|\mathbf{S}_1|/m_1^2 = |\mathbf{S}_2|/m_2^2 =
1$.  In these plots the initial orbital frequency is $\Omega_0
M=0.015$ which corresponds to a period $P=418M$. As can be seen, the
raw data (dotted lines) are dominated by the eccentricity at the
orbital frequency, while the residuals (solid lines) oscillate at
twice the orbital frequency, as expected. As far as the amplitude of
the residual oscillations is concerned, we see that it is
compatible with that of the predictions computed using
Eqs.~(\ref{rdotpred}), (\ref{omegadotpred}), even though the
agreement is not striking (red dashed lines, called non-precessing).
In fact, we find that the effectiveness of the removal procedure of
the homogeneous part is deeply affected by the value of
$\omega_\textrm{fit}$. Numerical studies have shown that differences
of a few percent on $\omega_\textrm{fit}$ can completely alter the
residuals. Tweaking by hand the value of $\omega_\textrm{fit}$ instead
of using the best fit value can lead to a much better agreement on the
amplitudes, at least for the first cycle. Note that in Fig.~\ref{PNprediction} no such ad hoc 
tweaking is used. Also, the fact that the
predictions quickly get out of phase with respect to the residuals is
mainly due to the assumption made in
Sec.~\ref{sec:ecc_conservative_SS} of keeping the spins constant or
non-precessing, that is using the evolution of the triad
$\{\mathbf{\hat{n}},\boldsymbol{\hat{\lambda}},
\mathbf{\hat{L}_{\textrm{N}}}\}$ specified by
Eqs.~(\ref{evn})--(\ref{evL}).  By contrast better phase agreement can
be obtained numerically if we use the time-evolution of the spins and
of the reference triad (blue dashed lines, called precessing) or
analytically if we had considered a reference triad in which the
precession of the orbital plane and spins were taken into account.

Another interesting feature is the relative importance of the spin-induced 
oscillations with respect to the eccentricity induced by the initial 
conditions. Both types of oscillations are showing up in the raw data 
of $\dot{r}$, while in the case of $\dot{\Omega}$ we only see
the oscillations due to the initial conditions. This can be
explained by our analytical predictions. Using Eqs.~(\ref{Newtr}), 
(\ref{Newtw}), we write for the eccentricity
\begin{equation}
e^{\rm{NS}}_r=\frac{\lvert B_r\rvert}{\bar{r}\,\bar{\Omega}}\quad\textrm{or}\quad 
e^{\rm{NS}}_{\Omega}=\frac{\lvert B_{\Omega}\rvert}{2\bar{\Omega}^2},
\end{equation}
where $B_r$ and $B_{\Omega}$ are the amplitudes of the oscillations of
the homogeneous solutions. We want to keep distinct notations for
$\dot{r}$ and $\dot{\Omega}$, even though at Newtonian level
$e^{\rm{NS}}_r=e^{\rm{NS}}_{\Omega}=e$ and therefore $\lvert
B_r\rvert=\bar{r}\lvert B_{\Omega}\rvert/2\bar{\Omega}$ . If we now
call $C_r$ and $C_{\Omega}$ the SS amplitudes, namely
\begin{equation}
C_r=-\frac{\bar{\Omega}\,C}{M^2\,\bar{r}}\,,\quad \quad \quad C_{\Omega}=-\frac{\bar{\Omega}^2\,C}{M^2\,\bar{r}^2},
\end{equation}
we parametrize the spin-induced oscillations in terms of a spin-induced ``eccentricity'',
\begin{equation}
e^{\rm{SS}}_r=\frac{\lvert C_r\rvert}{\bar{r}\,\bar{\Omega}}=\frac{\lvert C\rvert}{M^2\,\bar{r}^2}\,,\quad \quad \quad e^{\rm{SS}}_{\Omega}=\frac{\lvert C_{\Omega}\rvert}{2\bar{\Omega}^2}=\frac{\lvert C\rvert}{2M^2\,\bar{r}^2}.
\end{equation}
We have that the relative ratio is
\begin{equation}\label{eq:omgdotSS_vs_rdotSS}
\frac{ e^{\rm{SS}}_\Omega } { e^{\rm{SS}}_r } = \frac{1}{2},
\end{equation}
so it is expected that the significance of the spin-induced oscillations is smaller for $\dot{\Omega}$.

\section{Iterative eccentricity removal in presence of spins}
\label{sec:ecc_removal}

In the preceding sections, we showed that the PN Hamiltonian with
leading SS terms predicts oscillations with two distinct periods: the
orbital period, with amplitude and phase depending on initial
conditions that can genuinely be associated with orbital eccentricity;
and half the orbital period, independent of the initial conditions and
caused by spin-spin couplings.  We furthermore showed that these
spin-induced oscillations are suppressed in $\dot\Omega$ as compared
to $\dot r$ [see, e.g., Eq.~(\ref{eq:omgdotSS_vs_rdotSS})].

Our task is to find initial conditions that remove or at least
minimize the oscillations caused by eccentricity.  As in earlier work,
we shall begin with some trial initial conditions, evolve the binary
for about two orbits, analyze the motion of the black holes, and then
correct the initial conditions.  To exploit this suppression of
spin-induced oscillations in $\dot\Omega$, we will derive updating
formulae based on $\dot\Omega$.

\subsection{Updating formulae}
The basis for the updating formulae are the Newtonian expressions
 for distance $r$ and orbital frequency $\Omega$ (Eqs.~(\ref{r(e)}) and~(\ref{Newtw}))
\begin{eqnarray}
r_N(t) &=& \bar r\,\left[1-e\, \sin\phi(t)\right], \\
\Omega_N(t) &=& \bar \Omega\, \left[1 + 2e\, \sin\phi(t)
\right]\,, 
\end{eqnarray}
where $\phi(t)$ is the phase of the radial oscillations.  General
relativistic periastron advance will cause $\phi(t)$ to
deviate from the orbital phase.  Taking a time-derivative, we find
\begin{align}\label{eq:dotrN}
\dot{r}_N &=-\bar r\,e\,\omega\, \cos(\omega t+\phi_0),\\
\label{eq:dotOmegaN}
\dot\Omega_N&=2\bar\Omega\,e\,\omega \,\cos(\omega t+\phi_0),
\end{align}
with $\omega=(d\phi/dt)(0)$, $\phi_0=\phi(0)$.

Let us now consider a compact binary inspiral starting at $t=0$ at
initial separation $r_0$, with orbital frequency $\Omega_0$ and radial
velocity $\dot r_0$.  We take $\dot r(t)$ or $\dot\Omega(t)$ from a
general relativistic inspiral, and fit it with functional forms
\begin{align}\label{eq:dotrFit}
\dot{r}_{\rm NR}(t)&=S_r(t) + B_r\,\cos( \omega_r t + \phi_r ),\\
\label{eq:dotOmegaFit}
\dot{\Omega}_{\rm NR}(t)&=S_\Omega(t) + B_\Omega\, \cos( \omega_\Omega t+\phi_\Omega ).
\end{align}
The subscripts $r$ and $\Omega$ indicate whether the fit was performed
on $\dot r_{\rm NR}$ or $\dot\Omega_{\rm NR}$, and we will use a
bullet $\bullet$ in the subscript to represent either $r$ or $\Omega$.
The first part of each fit, $S_\bullet$, is a non-oscillatory function
that captures the radiation-reaction driven inspiral, whereas the
oscillatory piece captures the orbital eccentricity.  We neglect
spin-induced oscillations. The precise functional form of $S_\bullet$
is important, and sometimes it is advisable to include a quadratic
term $C t^2$ within the argument of the cosine.  We comment on these
considerations below in Sec.~\ref{sec:PracticalConsiderations}

Equation~(\ref{eq:dotrFit}) shows that at $t=0$, orbital eccentricity
contributes $\dot r_{\rm ecc,0}=B_r\cos\phi_r$ and $\ddot r_{\rm
  ecc,0}=-B_r\omega_r \sin\phi_r$ to the radial velocity and
acceleration.  Our goal is to now modify the initial data parameters
$\dot r_0$ and $\Omega_0$ such that $\dot r_{\rm ecc,0}$ and $\ddot
r_{\rm ecc,0}$ vanish.  The radial velocity is a free parameter of
the initial data, so $\dot r_{0,\rm new}=\dot r_0+\Delta \dot r$, where
\begin{equation}\label{eq:rdotUpdate-sep}
\Delta\dot r=-\dot r_{\rm ecc,0}= -B_r \cos\phi_r.
\end{equation}
To utilize our information about the radial acceleration $\ddot r_{\rm
  ecc,0}$ we recall that for the Newtonian Hamiltonian we have
\begin{equation}
\ddot{r}=\frac{\dot{p}_r}{\mu}
=-\frac{1}{\mu}\,\frac{\partial H_{\textrm{N}}}{\partial r}
=\frac{\mathbf{L}^2}{\mu^2 r^3}-\frac{M}{r^2}=r\,\Omega^2-\frac{M}{r^2}\,.
\end{equation}
A small change $\Omega_0\!\to\!\Omega_{0,new}\!=\!\Omega_0+\Delta\Omega$
therefore changes the radial acceleration by $\Delta\ddot r=2
r_0\Omega_0\Delta\Omega$.  This change cancels $\ddot r_{\rm ecc,0}$ when
\begin{equation}\label{eq:OmegaUpdate-sep}
\Delta\Omega=\frac{B_r\omega_r\sin\phi_r}{2r_0\Omega_0}.
\end{equation}
Equations~(\ref{eq:rdotUpdate-sep}) and~(\ref{eq:OmegaUpdate-sep}) are
one version of the updating formulae for the eccentricity
removal based on the separation coordinate. A sometimes more effective formula is presented below in 
Eq.~(\ref{eq:OmegaUpdate-sep-2}) which in earlier numerical work~\cite{Boyle2007,Chu2009,Lovelace2008} 
was applied to the {\em proper separation} between the horizons of the black holes.

A convenient way to derive updating formulae based on $\dot\Omega(t)$
begins by noting that the ratio of the amplitudes of oscillations in
Eqs.~(\ref{eq:dotrN}) and (\ref{eq:dotOmegaN}) is $-\bar
r/(2\bar\Omega)$.  Therefore, we obtain the desired updating formulas
by replacing $B_r$ with $-\bar r B_\Omega/(2\bar\Omega)$: 
\begin{eqnarray}
\label{eq:rdotUpdate-Omega}
\Delta\dot r&=& \frac{r_0 B_\Omega}{2\Omega_0}\,\cos\phi_\Omega,\\
\label{eq:OmegaUpdate-Omega}
\Delta\Omega&=&-\frac{B_\Omega\omega_\Omega}{4\Omega_0^2}\,\sin\phi_r.
\end{eqnarray}
[A sometimes more effective replacement for Eq.~(\ref{eq:OmegaUpdate-Omega}) 
is presented below in Eq.~(\ref{eq:OmegaUpdate-Omega-2}).]

\begin{figure}
\begin{center}
\includegraphics[width=0.43\textwidth]{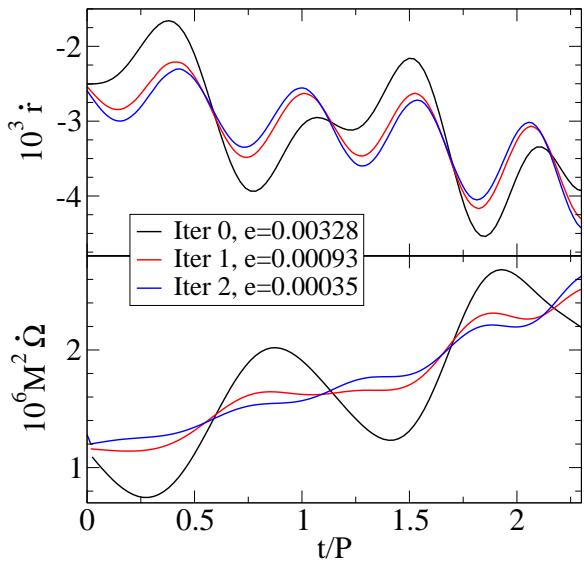}
\caption{Eccentricity removal based on $\dot\Omega$ applied to 
a {\bf PN model}. Shown is the initial orbital evolution,
and two iterations of eccentricity removal.
 Parameters of the 
  black-hole binary: mass-ratio $q=1$, spins of magnitude $\chi_1=\chi_2=0.8$
  with initial orientations $(\theta_1=0,\phi_1=0)$ and
  $(\theta_2=\pi/3,\phi_2=0)$ with respect to
  $\mathbf{\hat{L}_{\textrm{N\,0}}}$ and initial $M \Omega_0
  =0.0315$, that is an initial orbital period $P=200M$.  
  \label{PNalgo}}
\par\end{center}
\end{figure}

In Fig.~\ref{PNalgo}, we present three steps of this eccentricity
removal procedure. We use the PN-expanded Hamiltonian with
non-spinning terms up to 3PN order~\cite{DJS00,Buonanno06}, and
spinning terms up to 2PN order~\cite{Damour01c,Buonanno06}. The 
radiation-reaction effects are included through 2PN order as in 
Ref.~\cite{Buonanno06}.

We indicate in the legends the value of the eccentricity estimated
from the fitted amplitude of the oscillations, once the smooth
inspiral has been removed. Note that the initial orbital period is
about $200M$. At the $0^{\rm th}$ step the plots are showing the
evolution of the binary system with initial conditions determined
according to the procedure outlined in Ref.~\cite{Buonanno00}, 
leading to the presence of an initial eccentricity. From
the plots, we clearly see that we go from a situation dominated by
the homogeneous oscillations occurring at the average orbital
frequency (step 0) to the situation in which only spin-induced oscillations
occurring at twice the average orbital frequency are visible (step 2).

The configuration considered in Fig.~\ref{PNalgo} is close to merger, 
where the rapid evolution of the orbit makes it more difficult to
apply eccentricity removal.  In the next section we discuss 
how to improve the convergence of the iterative procedure.

\subsection{Practical considerations}
\label{sec:PracticalConsiderations}

Unfortunately, iterative eccentricity removal is sensitive to a
variety of effects which are not immediately obvious.  Without 
sufficient care, iterative eccentricity removal converges slowly, or
not at all.  In this section, we describe important details for the
effective and practical application of the eccentricity removal, as
well as diagnostics that allow users to evaluate whether the
eccentricity removal proceeds optimally.

The fits in Eqs.~(\ref{eq:dotrFit}) and~(\ref{eq:dotOmegaFit}) are 
used to compute the values of the oscillating part
$B_\bullet\cos(\omega_\bullet t+\phi_\bullet)$ at $t=0$.  Therefore it
is crucial that the function $S_\bullet$ that is intended to fit the
inspiral portion does {\em not} fit this oscillatory piece.
Initially, we used a polynomial for $S_\bullet$, but sometimes,
especially for shorter fitting intervals, such a polynomial picks up a
contribution of the oscillatory piece and results in an unusable
fit. Therefore we have constructed more robust fitting functions that
cannot capture oscillations.  Our current preferred choice is
\begin{equation}\label{eq:Sform}
S_\Omega^{(k)} = \sum _{n=0}^{k-1} A_k (T_c - t)^{-11/8 - n/4},
\end{equation}
with free parameters $A_k$ and $T_c$.  The functional form and the
exponents are motivated by PN inspirals, and we keep either $k=1$ or
$k=2$ terms of this expansion (for $k=2$, we use the same $T_c$ in both terms).

Another crucial ingredient for a reliable fit is a suitably chosen
fitting interval.  This interval needs to cover enough oscillations to
break degeneracies among the fitting parameters. However, if it
becomes too long, the evolution in the inspiral part will be harder to
capture with $S_\bullet$ and the quality of the fit will deteriorate.
 
Finally, the fit is used to compute $\dot r_{\rm ecc, 0}$ and $\ddot
r_{\rm ecc, 0}$, which are quantities at $t=0$.  It is desirable that
the fitting interval starts as close to $t=0$ as possible to minimize
extrapolation from the fitting interval to $t=0$.  However, a
numerical evolution relaxes in its early stages due a quasi
steady-state, and features during this relaxation need to be excluded
from the fitting interval.

A good means to ensure a satisfactory fit is to perform several fits,
and ensure that the results are consistent.  We perform four distinct
fits to $\dot\Omega$, where we change the order $k=1,2$ of the
inspiral component, Eq.~(\ref{eq:Sform}), and where we change the
order of the polynomial within the cosine in
Eq.~(\ref{eq:dotOmegaFit}) between $l=1$ as shown in
Eq.~(\ref{eq:dotOmegaFit}) and $l=2$ (i.e. adding a quadratic term
$C_\bullet t^2$).  In addition, we vary the location and length of
the fitting interval and check that the updates $\delta\dot r$ and $\Delta\Omega$ are unaffected.  

\begin{figure}
\centerline{\includegraphics[width=0.95\linewidth]{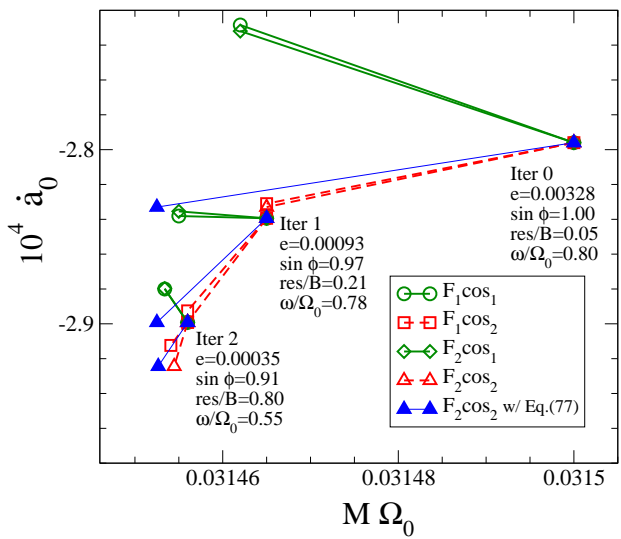}\hspace*{2em}}
\caption{\label{fig:EccRemoval} Visualization of the eccentricity
  removal performed in Fig.~\ref{PNalgo} in the $\Omega_0$--$\dot a_0$
  plane.  This plot summarizes a large amount of diagnostic
  information which can be utilized to ensure reliability of the
  eccentricity removal procedure (see main text).  The symbols $B$,
  $\phi$ and $\omega$ have a subscript $\Omega$ suppressed for
  clarity. }
\end{figure}

Figure~\ref{fig:EccRemoval} demonstrates a useful way to visualize and
assess the effectiveness of iterative eccentricity removal.  This plot
shows the plane of the initial-data parameters $\Omega_0, \dot a_0$,
with $\dot{a}_0 = \dot{r}_0/r_0$. The solid symbols correspond to the
three PN evolutions shown in Fig.~\ref{PNalgo}.  The lines emanating
from each of the symbols denote different fits based on this
particular evolution.  The different fits are denoted $F_k\,\cos_l$,
where $k=1,2$ denotes the fitting order in $S_\bullet$ and $l$ denotes
the order of the polynomial inside the cosine.  Each of these lines
ends at the predicted improved parameters $\Omega_{0,\rm new}, \dot
a_{0,\rm new}$.  Clustering of these lines, and convergence of the
symbols indicates that eccentricity removal is proceeding well.  As
can be seen by comparing the solid dark green and dashed red curves,
the order $k$ of the fitting function for the smooth part
$S_\Omega$ has almost no impact on the
  updated parameters $\Omega_{0,\rm new}, \dot a_{0,\rm new}$ in this
case. However,
  using a quadratic polynomial inside the cosine ($l=2$) significantly
  improves the quality of the $\dot\Omega_0$--update.

Figure~\ref{fig:EccRemoval} can also be used to assess the
potential quality for different updating formulae.  While we have kept
the orbital frequency $\Omega_0$ separate from the eccentricity
oscillation frequency $\omega_\Omega$, for Newtonian orbits both
frequencies agree.  Therefore, our Newtonian motivation does not
provide a means to choose whether to include extra powers of
$\Omega_0/\omega_\Omega$.  Specifically, we could replace
Eqs.~(\ref{eq:OmegaUpdate-sep}) and~(\ref{eq:OmegaUpdate-Omega}) by 
either
\begin{equation}
\label{eq:OmegaUpdate-sep-2}
\Delta\Omega=-\frac{B_r}{2r_0}\,\sin\phi_r,
\end{equation}
or
\begin{equation}
\label{eq:OmegaUpdate-Omega-2}
\Delta\Omega=-\frac{B_\Omega}{4\Omega_0}\,\sin\phi_\Omega,
\end{equation}
for updates based on $\dot r(t)$ and $\dot\Omega(t)$,
  respectively.  The predictions of the updating formula
  Eq.~(\ref{eq:OmegaUpdate-sep-2}) are shown in
  Fig.~\ref{fig:EccRemoval} as filled blue triangles.  It is obvious
  that Eq.~(\ref{eq:OmegaUpdate-sep-2}) predicts an updated $\Omega_0$
  significantly closer to the final best value for $\Omega_0$, even
  when applied to Iter 0.  Therefore, to summarize the discussion in
  the preceding paragraphs, for most effective eccentricity removal
  we recommend the fit of the form $F_2{\rm cos}_2$ combined with
  Eq.~(\ref{eq:OmegaUpdate-sep-2}). 

Finally, we discuss several diagnostics that can help to
assess the quality of eccentricity removal, and which are included
next to each symbol in Fig.~\ref{fig:EccRemoval}.
The first diagnostic is the estimated eccentricity $e$,
which should be monotonically decaying. The second diagnostic is
$\sin\phi_\Omega$.  As can be seen from 
Eqs.~(\ref{eq:rdotUpdate-Omega}) and~(\ref{eq:OmegaUpdate-Omega}), the
angle $\phi_\Omega$ parametrizes the relative importance of the
$\Omega_0$ and $\dot a_0$ updates.  For $\sin\phi_\Omega\approx 1$,
the whole weight is carried by the $\Omega_0$ update.  This is the
case here, and indicates that the starting value for $\dot a_0$ was
already very good, and that the apparent inconsistent predictions for
$\dot a_{0,\rm new}$ will not have an adverse impact on the
eccentricity fitting procedure (note that all fits predict consistent
values for $\Omega_{0,\rm new}$). The third diagnostic is the ratio
of the root-mean-square residual of the fit, ${\rm res}$, to the amplitude
of the oscillatory part, $B_\Omega$.  When $\mbox{res}/B\ll 1$, then
$\dot\Omega$ has indeed the assumed form Eq.~(\ref{eq:dotOmegaFit}), a
prerequisite for eccentricity removal.  When $\mbox{res}/B\sim 1$, we
can no longer isolate the oscillatory piece, and eccentricity removal
ceases to be effective. The final diagnostic is the ratio of
frequencies of radial oscillations $\omega_\Omega$ to orbital
frequency $\Omega_0$.  For a good fit, $\omega_\Omega/\Omega_0$ should
be somewhat less than unity, where the deviation from unity is caused
by periastron advance.  For moderately small eccentricities, this
ratio should furthermore be independent of the precise value of
eccentricity.  This is indeed the case for ``Iter 0'' and ``Iter 1'',
but ``Iter 2'' results in a questionably small ratio, which
furthermore differs from the values for iterations 0 and 1.  Again, an
indication that we cannot further proceed with eccentricity removal.

\section{Application to fully numerical binary black-hole simulations}
\label{sec:ecc_NR}

We now apply the method outlined in Sec.~\ref{sec:ecc_removal} to reduce
the initial eccentricity of single-spin and double-spin precessing 
binary black-hole simulations. We compare the periodicity in the 
oscillations of the orbital frequency and the proper horizon separation 
to the PN predictions described in Sec.~\ref{sec:ecc_AR} and also 
to the periastron-advance results of Ref.~\cite{Mroue2010}. Finally, 
for one binary configuration, we also extract the $l=2$, $m=2$ mode of the gravitational waveform and 
investigate the presence of spin-induced oscillations in its phase and frequency.

\subsection{Numerical methods}

Binary black hole initial data is constructed using the conformal thin 
sandwich formalism~\cite{York1999,Pfeiffer2003b} and
quasi-equilibrium boundary
conditions~\cite{Cook2002,Cook2004,Caudill-etal:2006}, incorporating
radial velocity as described in Ref.~\cite{Pfeiffer-Brown-etal:2007}.
The resulting set of five nonlinear coupled elliptic equations is
solved with multi-domain pseudo-spectral techniques described
in Ref.~\cite{Pfeiffer2003}.  As in earlier work, we choose conformal
flatness and maximal slicing.  To obtain desired masses and spins, we
utilize a root-finding procedure to adjust freely specifiable
parameters in the initial data~\cite{Buchman-etal-in-prep}.

Thus, a binary black-hole initial data set is determined by the mass-ratio, the
spins of both black holes, and coordinate separation $d$ between the
coordinate centers of the black holes, orbital frequency $\Omega_0$,
and radial velocity $\dot r_0 = \dot a_0 d$, where $\dot a_0$ is the
dimensionless expansion factor.  For fixed $d$, eccentricity removal 
consists of finding values for $\Omega_0$ and $\dot a_0$ that result in 
sufficiently small eccentricity.

The constructed initial data are evolved with the Spectral Einstein
Code {\tt SpEC} \cite{SpECwebsite}.  This code evolves a first-order
representation~\cite{Lindblom2006} of the generalized harmonic
system~\cite{Friedrich1985,Garfinkle2002,Pretorius2005c} and includes
terms that damp away small constraint 
violations~\cite{Gundlach2005,Pretorius2005c,Lindblom2006}.  The
computational domain extends from excision boundaries located just
inside each apparent horizon to some large radius.  No boundary
conditions are needed or imposed at the excision boundaries, because
all characteristic fields of the system are outgoing (into the black
hole) there.  The boundary conditions on the outer
boundary~\cite{Lindblom2006,Rinne2006,Rinne2007} are designed to
prevent the influx of unphysical constraint
violations~\cite{Stewart1998,FriedrichNagy1999,Bardeen2002,Szilagyi2002,%
  Calabrese2003,Szilagyi2003,Kidder2005} and undesired incoming
gravitational radiation~\cite{Buchman2006,Buchman2007}, while allowing
the outgoing gravitational radiation to pass freely through the
boundary. Interdomain boundary conditions are enforced with a penalty
method~\cite{Gottlieb2001,Hesthaven2000}.

\subsection{Eccentricity removal based on orbital frequency: single-spin binary black hole}
\label{sec:master_run}

\begin{figure}
  \begin{center}
    \includegraphics[scale=0.5]{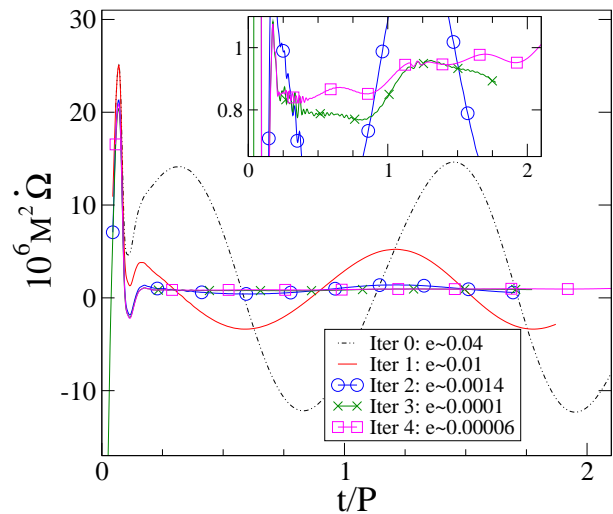}
    \caption{ \label{fig:EccRem-dOmegadt} {\bf Eccentricity removal
        based on time derivative of the orbital frequency
        $d\Omega/dt$}, applied to a single-spin {\em precessing}
      binary black hole with the same initial parameters as in
      Fig.~\ref{fig:EccRem-PropSep}.  Shown is $\dot{\Omega}$ vs. time
      in units of initial orbital period $P=442M$ for the initial run
      (based on PN parameters) and four eccentricity-removal
      iterations.  The amplitude of spin-induced oscillations is
      several orders of magnitude smaller than in
      Fig.~\ref{fig:EccRem-PropSep}, and becomes only visible in
      Iter~4.}
  \end{center}
\end{figure}

\begin{figure}
    \includegraphics[width=0.98\columnwidth]{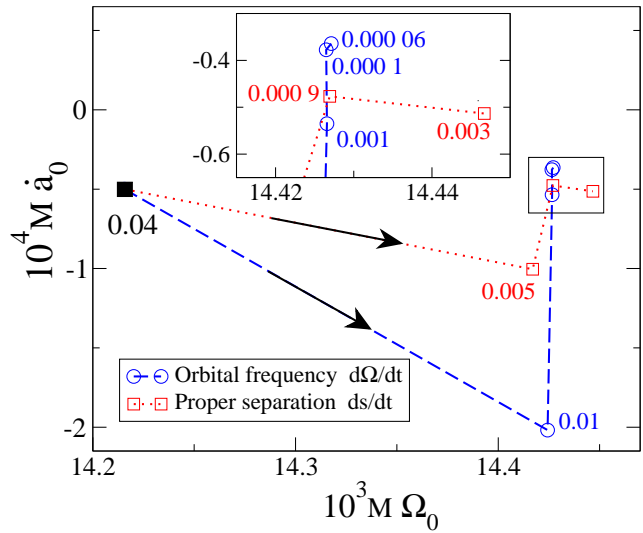}\hspace*{3em}
    \caption{{\bf Convergence of the eccentricity-removal procedures
        in the \boldmath$(\Omega_0 ,\dot{a}_0)$ plane.}  Blue circles:
      eccentricity removal sequence of Fig.~\ref{fig:EccRem-dOmegadt}.
      Red squares: eccentricity removal sequence shown in
      Fig.~\ref{fig:EccRem-PropSep}.  The number next to each symbol
      gives the eccentricity of the respective evolution.  The inset
      shows an enlargement of the boxed area.  The
      eccentricity-removal procedure based on the orbital frequency
      keeps converging until the fourth iteration, while the one based
      on the proper separation fails to converge any further beyond
      the second iteration.  }
    \label{fig:Convergence-adot-Vs-Omega}
\end{figure}

In this section we re-visit eccentricity removal for the configuration
considered in the introduction and Fig.~\ref{fig:EccRem-PropSep}.  The
binary has a mass-ratio of $m_1/m_2=1.5$, and only the larger black
hole carries spin, namely $\chi_1=0.5$ with initial spin direction in
the orbital plane pointing exactly away from the smaller black hole.
Note that spins in the orbital plane maximize spin-induced
oscillations [see, e.g., Eqs.~(\ref{partr}) and~(\ref{partomega})].  The
initial coordinate separation between the holes is $d\!=\!16M$,
orbital frequency $M\Omega_0\!=\!0.0142$, and $\dot a_0\!=\!-5\times
10^{-5}$.  These parameters were determined from the so-called 
TaylorT3 PN approximant for non-spinning binaries~\cite{Blanchet2006}.

\begin{figure*}
  \centerline{
    \includegraphics*[scale=0.5]{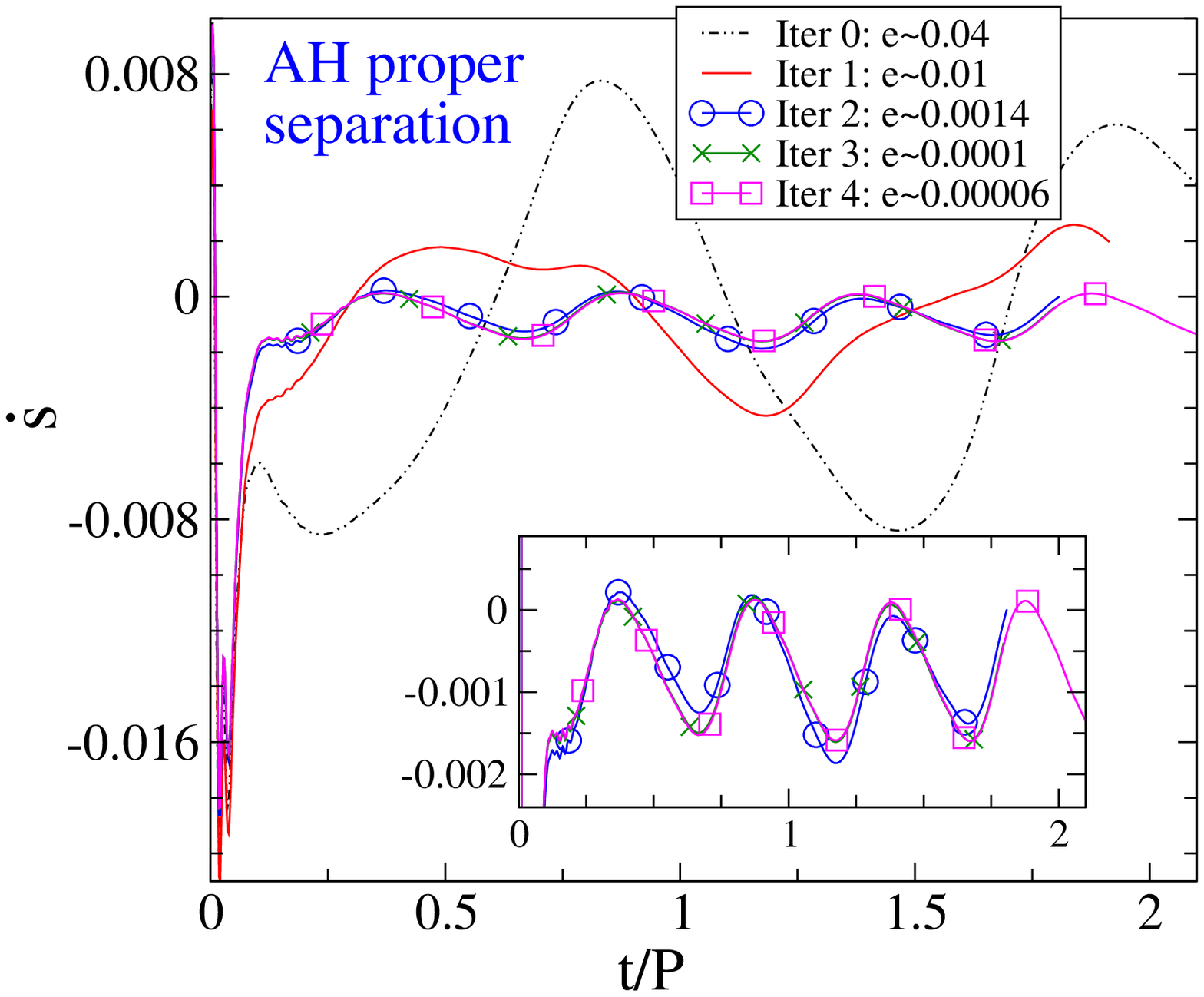}
    \hspace*{2em}
    \includegraphics*[scale=0.495]{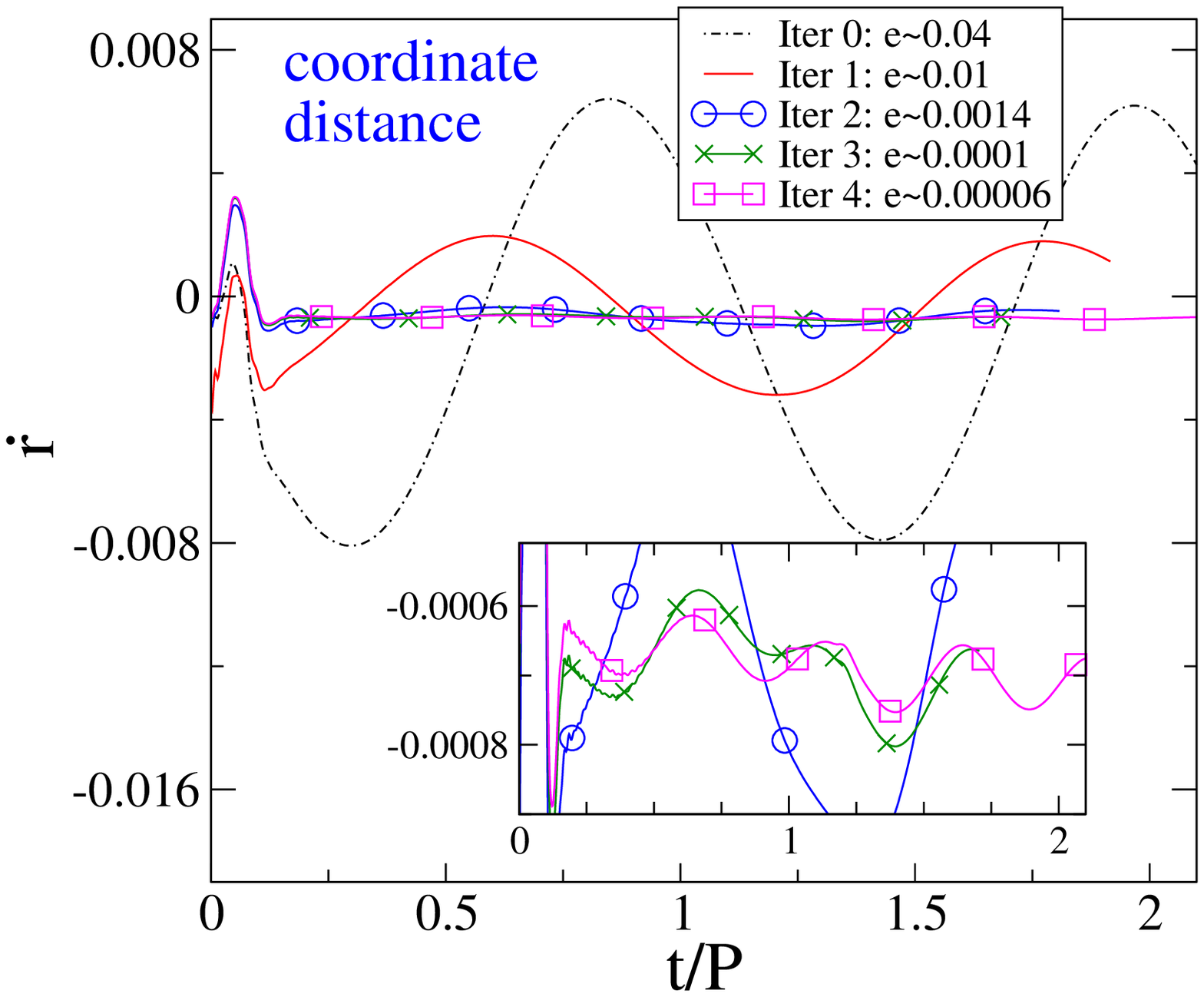}
 } 
 \caption{\label{fig:MasterRun-s-dsdt} {\bf Radial velocity} between
   the black holes for the same series of evolutions shown in
   Fig.~\ref{fig:EccRem-dOmegadt}.  {\bf Left:} Derivative of proper
   separation between the apparent horizons $\dot{s}$.  {\bf Right:}
   Derivative of coordinate distance between centers of the apparent
   horizons $\dot{r}$.  The time-axis is given in units of the initial
   orbital period $P=442M$.  Proper separation $\dot{s}$ exhibits large
   spin-induced oscillations, whereas $\dot{r}$ shows spin-induced
   oscillations of similarly small amplitude as in
   Fig.~\ref{fig:EccRem-dOmegadt}.  Note that the inset in the right panel is ten times more magnified than in the left panel.}
\end{figure*}

Orbital frequency, both in the initial data and in the subsequent
evolution, is defined by the coordinate motion of the center of the
apparent horizons.  Let $\mathbf {c}_{i}(t)$ be the coordinates of the
center of each black hole, and define their relative separation $\mathbf{
r}(t)= \mathbf{c}_1(t)-\mathbf{c}_2(t)$.  The instantaneous orbital frequency is
then computed as
\begin{equation}
\mathbf{\Omega} = \frac{\mathbf{r} \times{\mathbf{\dot r}}}{r^2},
\end{equation}
and $\Omega$ is defined as the magnitude of $\mathbf\Omega$. All these
calculations are performed using standard Euclidean vector calculus.

We start the first run using PN initial conditions for the orbital
frequency and radial velocity and evolve the binary black hole for
about two orbits. From the orbital frequency we measure an
eccentricity $e\sim 0.04$, and Eqs.~(\ref{eq:rdotUpdate-Omega})
and~(\ref{eq:OmegaUpdate-Omega}) give improved values for $\Omega_0$
and $\dot a_0$.  Evolution of the initial data computed
from these improved values is labeled ``Iter 1'' in
Fig.~\ref{fig:EccRem-dOmegadt}, and reduces the eccentricity to about
$0.01$. The same procedure is then repeated three more times. 
For Iter~0 to Iter~2, we
  exclude $t\lesssim 100M$ from the fit.  For Iter~3 the variations in
  $\dot{\Omega}$ are so small that numerical noise is dominant for about
  half an orbital period, and we exclude $t\lesssim 250M$ from the fit.
The final eccentricity in the orbital frequency is 
$e\sim6\times10^{-5}$.

In Fig.~\ref{fig:Convergence-adot-Vs-Omega}, we show how the
  initial orbital frequency $\Omega_0$ and the radial expansion factor
  $\dot a_0$ converge to the final values (minimal
  eccentricity).  The blue circles indicate the successive iterations
  for the successful eccentricity removal based on $\dot{\Omega}$ 
  (see Fig.~\ref{fig:EccRem-dOmegadt}).  Note that the parameters
  $(\Omega_0, \dot a_0)$ converge well for all iterations.  In
  contrast, the red squares denote the unsuccessful eccentricity
  removal based on proper separation $\dot{s}$
 (see Fig.~\ref{fig:EccRem-PropSep}).  Starting with the third
  iteration, the updated values of the orbital frequency and expansion
  radial coefficient move away from the line of minimum eccentricity,
  with an increase of eccentricity from $0.001$ to $0.003$.  All
  eccentricity estimates shown in this figure are computed from
  $\dot{\Omega}$, even when eccentricity removal is based on $\dot{s}$.
  This allows us to measure eccentricities $e<0.01$, which would not
  be possible from $\dot{s}$, because the latter is dominated by 
  large spin-induced oscillations.

The absence of spin-induced oscillations in
  Fig.~\ref{fig:EccRem-dOmegadt} is striking, especially when compared
  to Fig.~\ref{fig:EccRem-PropSep}.  Spin-induced oscillations are
  visible in Fig.~\ref{fig:EccRem-dOmegadt} only at $e<10^{-4}$.  For
  the runs with larger eccentricity (Iter~0--3), eccentricity--induced
  oscillations dominate with a period somewhat larger than $P$
  (somewhat larger because of periastron--advance~\cite{Mroue2010}).

  We shall now investigate spin-induced oscillations in the numerical-relativity 
  simulations in more detail.  First, by comparing the
  time-derivatives of orbital frequency $\dot{\Omega}$, proper
  separation between horizons $\dot{s}$, and coordinate separation
  between centers of apparent horizons $\dot{r}$.  Subsequently, by
  comparing their amplitude and frequency with PN predictions from
  Sec.~\ref{sec:ecc_AR}.

  Figure~\ref{fig:MasterRun-s-dsdt} shows time-derivatives of
  proper separation $\dot{s}$ and coordinate separation $\dot{r}$ for the
  evolutions shown in Fig.~\ref{fig:EccRem-dOmegadt}.  Spin-induced
  oscillations are already noticeable in $\dot{s}$ for Iter~1 with
  $e=0.01$.  These oscillations dominate for Iter 2--4, i.e. $e\leq
  0.0014$.  In contrast, the coordinate distance $\dot{r}$ is less
  susceptible to spin-induced oscillations.  In the right plot of
  Fig.~\ref{fig:MasterRun-s-dsdt}, spin-induced oscillations become
  apparent only for eccentricities of $\sim 10^{-4}$ or smaller.  The
  spin-induced oscillations in $\dot{r}$ are smaller by a factor of
  almost 20 than those in $\dot{s}$. 

  When comparing Iter 3 and Iter 4 between
  Fig.~\ref{fig:EccRem-dOmegadt} and the right panel of
  Fig.~\ref{fig:MasterRun-s-dsdt}, one notices that $\dot{\Omega}$ shows
  slightly less pronounced spin-induced oscillations.  That is
  consistent with the PN calculations, where
  Eq.~(\ref{eq:omgdotSS_vs_rdotSS}) predicted that spin-induced
  oscillations in $\dot{\Omega}$ should be suppressed by a factor of 2
  relative to those in $\dot{r}$.  When comparing Iter~4 between
  Fig.~\ref{fig:EccRem-dOmegadt} and the right plot of
  Fig.~\ref{fig:MasterRun-s-dsdt}, we find that the spin-induced
  oscillations in $\dot{\Omega}$ and $\dot{r}$ are {\em in phase}.  This
  is again consistent with the PN prediction, where the
  last terms of Eqs.~(\ref{deltadotr}) and~(\ref{deltadotomega}) have
  the same sign.  The phase of spin-induced oscillations in
  $\dot{\Omega}$ and $\dot{r}$ differs from the effect of orbital
  eccentricity: for an eccentric orbit, the orbital frequency is
  maximal when the separation is minimal, and therefore $\dot{\Omega}$
  and $\dot{r}$ are out of phase (see Iter~0 and Iter~1 of
  Figs.~\ref{fig:EccRem-dOmegadt} and~\ref{fig:MasterRun-s-dsdt}). 

\begin{figure}
  \includegraphics*[width=0.45\textwidth]{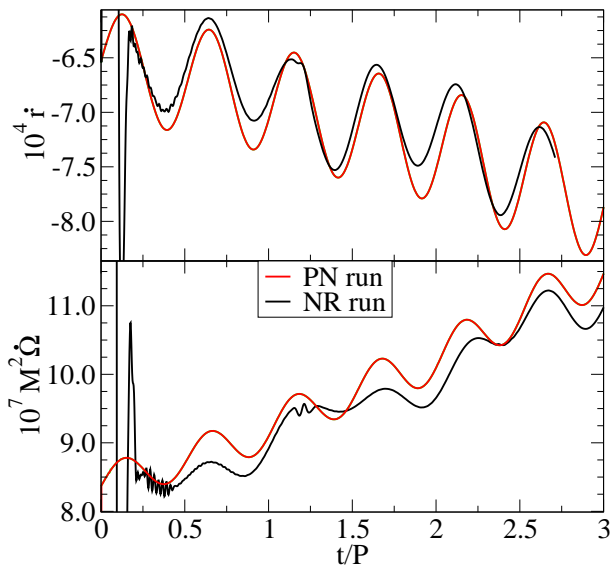}
  \caption{ \label{fig:NR-Vs-PN-after-rem} Spin-induced oscillations
    of the lowest-eccentricity numerical simulation (Iter~4 of Fig.~\ref{fig:EccRem-dOmegadt}) in comparison with 
    the PN calculations of Sec.~\ref{sec:ecc_conservative_SS}.}
\end{figure}

  We have shown in Sec.~\ref{sec:ecc_conservative_SS} that the
  PN Hamiltonian predicts spin-induced oscillations:
  Equations~(\ref{deltadotr})--(\ref{C}) contain an oscillatory component at
  twice the orbital frequency with amplitudes
\begin{align}
A_{\delta\dot r}&=\frac{\bar\Omega\, \mathbf S_{0\,\perp}^2}{2M^2\,\bar r},\\
A_{\delta\dot \Omega}&=\frac{\bar\Omega^2\, \mathbf S_{0\,\perp}^2}{2M^2\,\bar r^2}.
\end{align}
Figure~\ref{fig:NR-Vs-PN-after-rem} shows numerical data for the
lowest-eccentricity numerical simulation (Iter~4 from
Fig.~\ref{fig:EccRem-dOmegadt}). These numerical data are compared with the prediction of 
the PN equations. The PN calculation reproduces very accurately the amplitude of
spin-induced oscillations in the numerical-relativity simulation for $\dot{\Omega}$ and
$\dot{r}$. By contrast, the spin-induced oscillations in $\dot{s}$ are larger 
by a factor $\sim 20$ than those in $\dot{r}$. This can be due to 
deformations of the apparent horizons due to spin effects. 
Finally, we notice that a small amplitude oscillation
of the numerical data on the orbital time-scale is also visible,
corresponding to the small, but non-zero eccentricity $e=6\times
10^{-5}$ of the numerical simulation.

\subsection{Oscillations in the (2,2) mode of the gravitational wave} 
\label{sec:GW}

In Sec.~\ref{sec:master_run} we found spin-induced oscillations in 
the coordinate distance of the black holes and the orbital frequency, consistent
with PN predictions. We now investigate the gravitational radiation 
emitted by this binary.  Specifically, we extract the $l=2,m=2$ mode 
of the gravitational waveform in the inertial frame 
where the binaries are {\em initially} placed along the $x$-axis and the initial angular 
momentum is parallel to the $z$-axis. We compute phase and frequency 
for the waveforms extracted at extraction radii $R=130M$ and $R=220M$. 

Spin-induced oscillations represent a physical effect independent of
orbital eccentricity.  Nevertheless, the concept of eccentricity
estimators~\cite{Mroue2010} will be very useful when discussing
spin-induced oscillations, because it removes overall secular trends
(especially in the gravitational-wave phase), and because it makes it easy to relate
the amplitude of oscillations to an ``equivalent eccentricity.'' As
Ref.~\cite{Mroue2010}, we define $e_{\phi_{\rm GW}}(t)$ 
\begin{equation}
   \label{eq:ephase}
e_{\phi_{\rm GW}}(t)=\frac{\phi_{\rm NR}(t)-\phi_{\rm fit}(t)}{4 }\,, 
\end{equation}
where $\phi_{\rm NR}(t)$ is the gravitational-wave phase of the (2,2) mode 
and $ \phi_{\rm fit}(t)$ is the quasi-circular polynomial fit of the gravitational-wave phase 
[see Ref.~\cite{Mroue2010} for more details]. 
Similarly, using the gravitational-wave frequency of the (2,2) mode and its quasi-circular 
 polynomial fit as in Ref.~\cite{Mroue2010}, we define the eccentricity estimator 
$e_{\omega_{\rm GW}}(t)$ 
\begin{equation}
\label{eq:efrequency}
e_{\omega_{\rm GW}}(t)=\frac{\omega_{\rm NR}(t)-\omega_{\rm fit}(t)}{2 \omega_{\rm fit}(t) }\,.
\end{equation}

\begin{figure}
  \begin{center}
    \includegraphics*[width=0.45\textwidth]{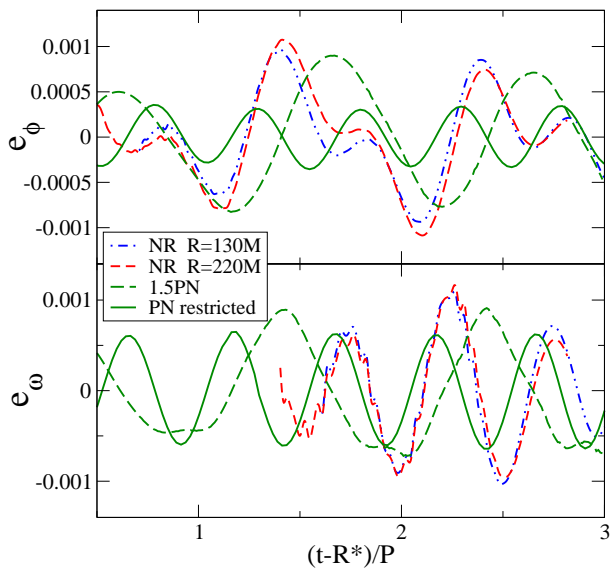}
    \caption{{\bf Eccentricity estimator for the gravitational-wave
        phase and frequency} for the final eccentricity-removal
      iteration of Fig.~\ref{fig:EccRem-dOmegadt}.  The upper panel
      shows the gravitational-wave--phase eccentricity estimator for
      two extraction radii versus the retarded time, the lower panel
      the eccentricity estimator computed from the gravitational wave
      frequency. In both panels we clearly see 
      oscillations at twice the orbital frequency. We also show the PN
      result for the restricted waveform.
    \label{fig:WavePhase}}
  \end{center}
\end{figure}

In Fig.~\ref{fig:WavePhase}, we plot the eccentricity estimators
$e_{\phi_{\rm GW}}(t)$ (upper panel) and $e_{\omega_{\rm GW}}(t)$ (lower panel) 
for the two extraction radii $R=130M$ and $R=220M$ versus the retarded time
$t-R^*$, where $R^*$ is the tortoise-coordinate radius defined as 
\begin{equation}
  R^*\equiv R + 2 M\, \ln \left( \frac{R}{2 M}-1\right) \ , 
\end{equation}
where $M=1$ is the total mass of the initial data. 
Quite interestingly, the plots show oscillations happening at
twice the orbital frequency.  The magnitude of the oscillations in
$e_{\phi_{\rm GW}}(t)$ or $e_{\omega_{\rm GW}}(t)$ is $\sim 10^{-3}$, although the
eccentricity in the orbital frequency has been reduced to $\sim
6\times 10^{-5}$ (see Fig.~\ref{fig:EccRem-dOmegadt}).  We note that
the amplitude of the oscillations at twice the 
orbital frequency does not depend on the
extraction radius, suggesting that they are gauge invariant.

We also compare these numerical result with what is predicted by the
PN model. For the orbital evolution we use the model Hamiltonian
(\ref{modelH}), where SO and SS interactions are included through 2PN
order, nonspinning effects through 3PN order, and radiation-reaction
effects through 2PN order. As to the analytical model, 
we employ the waveform derived in Ref.~\cite{Arun:2008kb}, where the precession of the orbital plane and
the spins of the black holes are taken into account through 1.5PN
order. In particular, we compute the estimators $e_{\phi_{\rm GW}}$ and
$e_{\omega_{\rm GW}}$ using Eqs.~(4.15), (4.16a) in Ref.~\cite{Arun:2008kb} for the $h_{22}$ mode with the amplitude
computed at lowest order in $v/c$ (restricted waveform)\footnote{Since the numerical-relativity  
(2,2) mode is computed using as $z$-axis the direction perpendicular to the 
orbital plane, we apply the Wigner rotation to the restricted $h_{22}$ of 
Ref.~\cite{Arun:2008kb} and keep only the lowest-order term in $v/c$.}. 
This means that the precession of the orbital plane is considered only in the
gravitational-wave phase, but not in the amplitude.

  Before computing the PN eccentricity estimators, we perform an
  alignment between the phase of the PN $h_{22}$ and the phase of the
  numerical-relativity $\Psi_4$. To do that, we follow the procedure
  outlined in Sec. IIIA of Ref.~\cite{Pan:2009wj}. This alignment is
  obtained over a time window of $1000M$ (in the range
  $500M<t<1500M$), and it amounts to a time-shift and a global offset
  in the PN phase. The result is shown as solid black curves in
  Fig.~\ref{fig:WavePhase}. We see a qualitative agreement between
  numerical-relativity results and the restricted PN model for the 
  oscillations at twice the orbital frequency in $e_{\omega_{\rm GW}}$ and $e_{\phi_{\rm
      GW}}$. However, the numerical-relativity $e_{\phi_{\rm GW}}$
  also shows oscillations at the orbital frequency
  which are absent in the restricted PN waveform. We find that
  oscillations at the orbital frequency can be generated in the PN
  model of Ref.~\cite{Arun:2008kb} if we included higher order PN
  corrections in the amplitude of the (2,2) mode [see Eq.~(4.16a) in
  Ref.~\cite{Arun:2008kb}]. Such oscillations cannot be iterated away 
  by our procedure, even in principle, since the removal algorithm concerns the orbital 
  dynamics and they rather appear as a physical effect of the waveform. The upper panel of
  Fig.~\ref{fig:WavePhase} shows comparatively large oscillations at
  period $\sim P$; because the PN model predicts modes at this
  frequency, these oscillations cannot be used to compute orbital
  eccentricity.  A further analysis of the inclusion of higher-order
  PN corrections is warranted. We prefer to postpone such an
  analysis to be able to test against a larger sample of
  numerical--relativity waveforms.

\subsection{Eccentricity-removal for generic binary black holes} 

\label{sec:other_rus}

\begin{figure}
  \includegraphics*[width=0.45\textwidth]{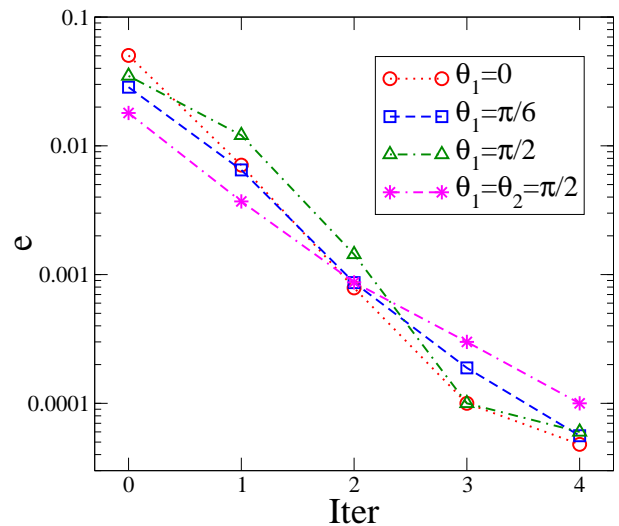}
    \caption{{\bf Eccentricity removal for different
          spin--configurations.} We illustrate of how the eccentricity
      is reduced to very low values when the iterations are applied to
      the orbital frequency.  Shown are three configurations with 
      $S_1/M_1^2\!=\!0.5, S_2\!=\!0$ and different spin directions 
      $\theta_1$, and one configuration with $S_1/M_1^2\!=\!S_2/M_2^2\!=\!0.5$, 
      with initially two orthogonal spins both tangent to the orbital plane.  
      For all cases, the mass-ratio is $m_1/m_2\!=\!1.5$.  The run shown in green 
      triangles was discussed in detail in Figs.~\ref{fig:EccRem-dOmegadt} 
      to~\ref{fig:WavePhase}. 
      \label{fig:q1.5d16eN}  
    }
\end{figure}

In the previous sections we studied our new eccentricity-removal
procedure in detail for one test-case: a binary with only one non-zero
spin, and with mass-ratio $m_1/m_2 = 1.5$.  We now test the procedure
for other binary configurations with the same mass ratio.  We consider
two further configurations where only the large black hole carries
spin, parametrized by the angle $\theta_1$ between the orbital angular
momentum and the spin axis of the first black hole.  In the previous
sections, we considered $\theta_1=\pi/2$, and now we extend to
$\theta_1=0,\pi/6$, and $\phi_1=0$.  The first of these new cases is
non-precessing and verifies that eccentricity removal based on
$\dot{\Omega}$ works effectively for non-precessing systems. We also
consider a binary where both black holes carry spin, with initial
spin-direction in the orbital plane ($\theta_1=\theta_2=\pi/2$),
$\mathbf S_1$ parallel to the initial separation vector between the
black holes, and $\mathbf S_2$ normal to the separation vector.  (All
spinning black holes have dimensionless spin-magnitude of $0.5$.)
Figure~\ref{fig:q1.5d16eN} demonstrates the effectiveness of the
eccentricity removal procedure based on $\dot{\Omega}$ and
Eqs.~(\ref{eq:rdotUpdate-Omega}) and~(\ref{eq:OmegaUpdate-Omega}).
In all cases, the eccentricity is reduced to less than $10^{-4}$ in four iterations.

The number of required eccentricity removal iterations depends
  on the quality of the guess for $\Omega_0$ and $\dot a_0$ for the
  first iteration.  Once eccentricity removal has been performed for
  several different configurations, we expect to be able to
  interpolate between configurations, to improve the initial guess
  substantially.

\section{Conclusions}
\label{sec:conclusions}

The removal of the initial spurious orbital eccentricity in binary
black-hole simulations is quite challenging, and it becomes more
complicated in the presence of spins. As predicted by PN
theory, and worked out in Sec.~\ref{sec:ecc_AR}, spin-spin interactions (notably
${S}_1\,{S}_1$, ${S}_2\,{S}_2$ and ${S}_1\,{S}_2$ terms) and
precession induce oscillations in the binary radial separation and
orbital frequency. These {\em spin-induced oscillations} are also
present in the gravitational radiation emitted by the binary, and 
their frequency is close to twice the average orbital frequency.  In
Sec.~\ref{sec:ecc_NR} we confirm the presence of spin-induced oscillations
in fully numerical simulations of black hole binaries.  The agreement
between the numerical simulations and PN calculations is
as good as can be expected given the low order of the PN
calculations, and the differing coordinate gauges.

  Spin-induced oscillations can be distinguished from
  oscillations caused by orbital eccentricity by their characteristic
  frequencies.  Moreover, the amplitude of spin-induced oscillations is
  quite small, so that it becomes visible only at small
  eccentricities, as can be seen from
  Eq.~(\ref{eq:spin-induced--ecc-bound}): At separations relevant for
  numerical simulations, spin-induced oscillations dominate orbital
  eccentricity only for $e\lesssim 0.001$, even for maximal spins in
  the least favorable orientation (parallel spins in the orbital
  plane).  The amplitude of spin-induced oscillations is proportional
  to $S_{0,\perp}^2$, so that for spin $S/M^2\sim 0.5$ as considered
  here, spin-induced oscillations become visible at orbital
  eccentricity $e\sim 10^{-4}$.

  Spin-induced oscillations affect the orbital frequency
  derivative $\dot{\Omega}$ less than the radial velocity $\dot{r}$.
  Therefore, we develop in
Sec.~\ref{sec:ecc_removal} a new eccentricity-removal procedure based
on the derivative of the orbital frequency, and apply it to
PN inspirals.  Subsequently, we successfully apply the
eccentricity removal procedure to fully numerical binary black hole
evolutions to achieve eccentricities smaller than $10^{-4}$.  At this
residual eccentricity, spin-induced oscillations begin to dominate
over orbital eccentricity oscillations, as expected from our
PN calculations.  In Sec.~\ref{sec:other_rus}, we tested
the new eccentricity-removal procedure on fully numerical binary black
hole simulations for several different spin configurations.

  The computational cost for eccentricity reduction depends on
  the number of eccentricity removal iterations.  Great care is
  necessary when performing the fitting, in order to achieve a large
  reduction in eccentricity per iteration.
  Section~\ref{sec:PracticalConsiderations} gives guidance to improve
  these fits.  With a good initial guess of $\Omega_0, \dot{a}_0$ for
  the 0-th iteration, one can start eccentricity removal from an already
  small initial eccentricity.  As the number of data-points increases,
  we expect to be able to compute a better initial guess by
  interpolating between already computed low-eccentricity binary black-hole 
  configurations.

  Perhaps surprising, our present study indicates that
  eccentricity removal should {\em not} be based on the proper
  separation between the apparent horizons.  These new findings
  supersede the practice of earlier
  papers~\cite{Chu2009,Lovelace2008,Boyle2007} to base eccentricity
  removal on proper separation rather than coordinate separation to
  take advantage of reduced numerical noise.  As apparent in
  Figs.~\ref{fig:MasterRun-s-dsdt} and~\ref{fig:NR-Vs-PN-after-rem},
  spin-induced oscillations in $\dot{s}$ are about 15 times larger than
  in $\dot{r}$.  Therefore, eccentricity-removal based on $\dot{s}$ will
  fail at $\sim 15$ times larger eccentricity than using $\dot{r}$, and
  at $\sim 30$ times larger eccentricity than for $\dot{\Omega}$.  A
  likely cause for the unsatisfactory behavior of $\dot{s}$ lies in the
  deformation of the apparent horizons due to spin.  For spins with a
  component in the orbital plane, the ``bulge'' of the apparent
  horizon rotates through the line connecting the two black holes as
  the black holes orbit each other. Earlier work~\cite{Chu2009,Lovelace2008,Boyle2007} considered spins
  aligned with the orbital angular momentum, where this effect is
  absent; in those cases use of $\dot{s}$ was in order --- but for
  precessing binaries, use of $\dot{s}$ is not advisable.
  
  Even when the orbital frequency indicates $e < 10^{-4}$ for a fully
  numerical binary black-hole simulation, the extracted (2,2)
    mode of the gravitational radiation still shows oscillations at
  the orbital frequency in the wave phase with amplitude $\sim
  10^{-3}$. While future work is necessary for a detailed
    understanding, the PN model predicts oscillations in the GW at the
    orbital frequency, and therefore, one cannot use the gravitational
    waveforms to estimate orbital eccentricity for precessing
    binaries.  The wave phase and frequency of the NR simulation also shows oscillations
  at twice the orbital frequency with amplitude $\sim 3\times10^{-4}$
  which are qualitatively reproduced by the restricted PN model of
    Ref.~\cite{Arun:2008kb}.  We postpone the study of the details of
  these features in the gravitational waveform to future work.  Quite
  interestingly, it proves, for this particular binary configuration
  in which only one hole spins, that those spin-induced oscillations
  are a direct consequence of monopole-quadrupole
  interactions~\cite{Poisson:1997ha,Damour01c,Gergely:2002fd,Racine:2008qv,Racine2008}.

All fully relativistic simulations presented here were performed using
generalized harmonic coordinates. It would be very interesting to
perform a similar study within the moving-puncture BSSN approach to
investigate whether our conclusions are applicable in other gauges.

\begin{acknowledgments}
  We thank Guillaume Faye, Geoffrey Lovelace, Evan Ochsner and Yi Pan for useful
  interactions. The presented binary black hole evolutions were
  performed using the Spectral Einstein Code {\tt
  SpEC}~\cite{SpECwebsite}.  A.B. and A.T. acknowledge support from
  NSF Grant PHY-0903631.  A.B. also acknowledges support from NASA
  grant NNX09AI81G. A.T. also acknowledges support from the Maryland Center for 
Fundamental Physics. L.K. is supported by NSF Grants No. PHY-0969111
  and No. PHY-1005426 at Cornell; and by NASA Grant No. NNX09AF96G at
  Cornell.  H.P. gratefully acknowledges support from the NSERC of
  Canada, from Canada Research Chairs Program, and from the Canadian
  Institute for Advanced Research.  Computations were performed on the
  GPC supercomputer at the SciNet HPC Consortium. SciNet is funded by:
  the Canada Foundation for Innovation under the auspices of Compute
  Canada; the Government of Ontario; Ontario Research Fund - Research
  Excellence; and the University of Toronto.
\end{acknowledgments}

\bibliography{References/References}
\end{document}